%% file: src.tex
\documentclass[journal,10pt,draftclsnofoot,onecolumn]{IEEEtran}

\usepackage[utf8]{inputenc}
\usepackage[T1]{fontenc}
\usepackage{url}
\usepackage{ifthen}
\usepackage{cite}
\usepackage[cmex10]{amsmath} 
\usepackage{bbm}

\usepackage{mathtools,amssymb}
\mathtoolsset{showonlyrefs=true}

\newtheorem{thm}{\magentanote{Theorem}}
\newtheorem{cor}{Corollary}
\newtheorem{lemma}{\greennote{Lemma}}
\newtheorem{assume}{\bluenote{Assumption}}
\newtheorem{proposition}{Proposition}

\input{macro.tex}

\begin{document}
%
\title{Improved MDL Estimators Using Fiber Bundle of Local Exponential Families for Non-exponential Families}
%
%
%

\author{Kohei~Miyamoto,~\IEEEmembership{Non~Member,~IEEE,}
    Andrew~R.~Barron,~\IEEEmembership{Senior~Member,~IEEE,}
    and Jun'ichi~Takeuchi,~\IEEEmembership{Member,~IEEE}
\thanks{K. Miyamoto is with the Cybersecurity Laboratory, Cybersecurity Research Institute,
National Institute of Information and Communications Technology,
4-2-1 Nukui-Kitamachi, Koganei, Tokyo 184-8795, Japn}%
\thanks{A.~R.~Barron is with the Department of Statistics and Data Science, Yale University, 
Kline Tower 219 Prospect Street, New Haven, CT 06511, USA.}
\thanks{J. Takeuchi is with
the Department of Electrical Engineering and Computer Science, Kyushu University,
744 Motooka, Nishi-ku, Fukuoka, Fukuoka 819-0395, Japan }
\thanks{This research was partially supported by JSPS KAKENHI Grant Numbers JP18H03291 and JP23H05492. This paper was presented in part at the 2019 IEEE International Symposium on Information Theory, Paris, France, July 7-12, 2019.}}

%
%

\markboth{IEEE Transactions on Information Theory}%
{Shell \MakeLowercase{\textit{et al.}}: Bare Demo of IEEEtran.cls for Journals}
%



\maketitle

\begin{abstract}
  Minimum Description Length (MDL) estimators,
  using two-part codes for universal coding, are analyzed.
  For general parametric families under certain regularity conditions,
  we introduce a two-part code whose regret is close to the minimax regret,
  where regret of a code with respect to a target family $\Mset$
  is the difference between
  the code length of the code and the ideal code length achieved
  by an element in $\Mset$.
  This is a generalization of the result for exponential families by Gr\"unwald.
  Our code is constructed by using
  an augmented structure of $\Mset$
  with a bundle of local exponential families
  for data description, which is not needed for exponential families.
  This result gives a tight upper bound on risk and loss
  of the MDL estimators
  based on the theory introduced by Barron and Cover in 1991.
  Further, we show that we can apply the result to mixture families,
  which are a typical example of non-exponential families.
\end{abstract}


%
\IEEEpeerreviewmaketitle

\section{Introduction}
The density estimation problem is investigated.
Given a data string $x^n$ drawn from an unknown $p^*$,
we want to estimate $p^*$.
We consider density estimation for parametric families,
based on two-part codes in accordance with an instance of
the minimum description length principle \cite{Rissanen_1978}.
A two part code is used to compress the data string $x^n$,
where the first part is an encoding of a density 
and the second part is an encoding of
the data by the Shannon code using  that density.
In a parametric family $\Mset$ 
$=\{p_\theta : \theta \in \Theta \subseteq \Rset^K\} $, a density is described by an encoding of its parameter values.
We choose the density which gives the minimum code length of the two part code.
In this way we can design an estimator, which is called as an MDL estimator.
We examine the information theoretic regret and the statistical risk of these MDL estimators.
The statistical risk of an estimator is
the expectation of the loss of the estimator.
We employ the R\'enyi divergence \cite{Renyi} as the loss function.
It is known that if the true density is in the parametric family, then 
an upper bound of the risk of an MDL estimator is given by
the expected regret of the two part code which induces the MDL estimator \cite{barron1985logically,Grunwald}.
Hence,
if we design a two part code which has small expected regret, then 
the MDL estimator enjoys a corresponding risk bound.

The regret of a code is
defined for each $x^n$ by the difference of its code length
from the minimum of the Shannon code lengths for densities in the family, 
which corresponds to the maximum likelihood value of the parameters.
The minimax value of the regret provides a criterion of optimality.
It is known that the minimax optimal code is the normalized maximum likelihood (NML) codes with respect to $\Mset$ \cite{Shtarkov},
which achieves the uniform regret for all $x^n$.
The code length of the NML code is called the stochastic complexity of the data with respect to $\Mset$.

In this paper, we assume $p^*$ belongs to the $K$ dimensional parametric model $\Mset$.
We try to obtain the risk bound of MDL estimators
through evaluating their regret bounds with respect to $\Mset$.
It is known that when $\Mset$ is an exponential family,
by using a quantization of $\Mset$ with its precision and shape 
determined by the Fisher information matrix $J(\theta)$,
a two part code can be constructed which enjoys the regret close to the minimax regret with respect to $\Mset$.
In fact, Gr{\"u}nwald \cite{Grunwald} gave an upper bound on
the regret of such a code in the following form:
\begin{align}
  \label{intro_Grunwald_regret}
\frac{K}{2} \log n + \log \int_{\Theta} \abs{J(\theta)}^{1/2} d\theta + \frac{Ka^2}{8} - K \log a + o(1),
\end{align}
where $o(1)$ is a value which converges to $0$ when $n$  
goes to infinity
and $a$, which is an arbitrary positive constant, is a parameter of Gr{\"u}nwald's coding scheme. Note that the upper bound is minimized at $a=2$.
Here the symbol $\log$ denotes natural logarithm and we measure code length with bit through this paper.
This upper bound is close to the asymptotic evaluation of the minimax regret given in 
\cite{Rissanen_1996,Takeuchi_Barron_1998}
under certain regularity conditions, which is
\begin{align}
  \label{asymptotic_minimax_regret}
  \frac{K}{2} \log \frac{n}{2 \pi} + \log \int_{\Theta} \abs{J(\theta)}^{1/2} d\theta + o(1).
\end{align}
This means that the two part code 
for an exponential family
is close to the optimal code in terms of regret.
To obtain a risk bound of a two-part MDL estimator,
we may slightly modify the code.
In concrete,
we simply multiple the model description length by a constant $\alpha > 1$.
For $\alpha \ge 1$ (allowing for no modification with $\alpha = 1$)
the regret of the modified code has an upper bound of the following form.
\begin{align}
  \label{intro_Grunwald_regret_with_alpha}
  \alpha \pair[\Big]{\frac{K}{2} \log n + \log \int_{\Theta} \abs{J(\theta)}^{1/2} d\theta - K \log a} + \frac{Ka^2}{8}  + o(1).
\end{align}
This regret bound is also close to the minimax regret, for $\alpha$ near $1$.
This leads to a suitable upper bound for the risk of the MDL estimator.

The asymptotic regret bound \eqref{intro_Grunwald_regret} is shown only for exponential families.
It would be desirable
to establish two part codes for non-exponential families with the same asymptotic regret bound. 


One could consider
the same quantization as for exponential families.
However, that would suffer an increase of the regret caused by the difference between Fisher information and empirical Fisher information,
which is related to the fact that
the model is not an exponential family.

To overcome this difficulty, we use
a model enlargement technique using both Fisher information
and empirical Fisher information, which
corresponds to a fiber bundle of local exponential families \cite{Amari_Nagaoka}.
We design
a new two part code using the enlarged model which has the same asymptotics
as the bound for the exponential family case.

Analogous concerns have arised previously \cite{Takeuchi_Barron_1998}
in the context of identifying the asymptotics of minimax regret \eqref{asymptotic_minimax_regret}
using Bayes codes, based on the prior of Jeffreys
(with boundary modification). This Jeffreys prior works for the minimax regret for exponential families,
but to handle smooth non-exponential families as in \cite{Takeuchi_Barron_1998,Takeuchi_Barron_2013,Takeuchi_Barron,Takeuchi_Barron_2014}
a slightly enlarged 
family is used with local exponential tilting (a fiber bundle) based 
on the discrepancy between Fisher information and 
empirical Fisher information. The present paper establishes
that a similar fiber bundle allows two part regret conclusion
do similarly carry over from the exponential family case to the non-exponential case.

The regret bound \eqref{intro_Grunwald_regret_with_alpha} 
leads to the risk bound of the MDL estimators,
owing to \cite{Barron_Cover,Chatterjee_Barron,Li_estimation_mixture,Zhang_convergence_MDL,Barron_etal2008,Grunwald},
of the form of
\begin{align}
  \E_{X^n \sim p^*} \dbar_{\lambda}(p^* \| \ddot{p}_{X^n}) 
  \leq
  \frac{\alpha}{n}
  \pair[\Big]{
    \frac{K}{2} \log n + \log \int_{\Theta} \abs{J(\theta)}^{1/2} d\theta  - K \log a
  }
  + \frac{K a^2}{8n}
  + o\Bigl(\frac{1}{n}\Bigr),
  \label{intro_risk_bound}
\end{align}
where
$\ddot{p}_{X^n}$ denotes the MDL estimate obtained by the data $X^n$
and
$\dbar_{\lambda}$ is the R\'enyi divergence of order $\lambda$ (we assume $0< \lambda < 1$)
which is used as the loss function for estimators.
Note that $\alpha$ must satisfy $\alpha \ge 1/(1-\lambda)$.

We also show that
the mixture families
satisfy the assumptions we need to construct the improved two part code.
Then, the MDL estimator induced by the two part code for mixture families
has a nice risk guarantee.
A mixture family is defined as the family of convex combinations of known probability densities.
These mixture families provide typical examples of non-exponential families.

The primary scope of this paper is estimation in a single parametric family,
but we can use the result of this paper
to solve the estimation problem for multiple families, which includes a role for the model selection.
In fact,
by introducing a three part code 
where the first part is to encode which family and
then a two part code within each family, 
we can construct a code based on the multiple models.
Then we obtain an MDL estimator based on the multiple models.
The risk of such estimator is bounded by using the upper bound of the risk of the MDL estimator on each model
which is provided by \eqref{intro_risk_bound}
with an additional term for the description of the models.
Such estimator allows us
to select a model as well as to estimate a density
with a guarantee of the risk.
This risk bound will exhibits the trade-off between complexity of the models 
and the accuracy of their approximation.

\section{Two part codes and MDL estimators}
In this section, we review the notion of two part code
and MDL estimator.
Then, we review some results on the risk of MDL estimators.

An important result is the one given by
\cite{Barron_Cover,Chatterjee_Barron,Li_estimation_mixture,Zhang_convergence_MDL,Barron_etal2008,Grunwald}.
It states that
the MDL estimator induced by a two part code 
has an upper bound on its statistical risk which equals the redundancy of the code,
that is, 
when the redundancy of the two part code is small, then
the upper bound on the risk of the MDL estimator is small.
Hence, to obtain a good MDL estimator,
we should construct a two part code which enjoys a small redundancy.

We note that the original form of risk bound in \cite{Barron_Cover}
is given by the index of resolvability, which is an upper bound on redundancy
and is related to the trade-off between complexity of the models and 
the accuracy of their approximation.

Let $\Xset$ be a measurable set,
$\Mset = \{p_\theta : \theta \in \Theta  \}$
a parametric model of probability densities over $\Xset$,
and
$\set{\ddotPset_n}$ a sequence of countable sets of probability densities over $\Xset$.
Usually $\ddotPset_n$ are assumed to be subsets of $\Mset$,
but it is not always assumed.
In this paper,
we let $x^n$ denote a string $x_1\ldots x_n \in \Xset^n$.
For a probability density $p$ over $\Xset$,
we let $p(x^n)=\prod_{t=1}^n p(x_t)$, i.e.\
we assume $x^n$ is i.i.d.
Further, for each $n \in \set{1, 2, \ldots}$,
let $L_n$ be a code length function satisfying the Kraft's inequality on $\ddotPset_n$.
For a fixed $\alpha \geq 1$,
we define a code length function $\Ltp$ on $\Xset^n$ as
\begin{align}
  \label{definition_L2p}
  \Ltp(x^n) = \min_{p \in \ddotPset_n} \pair[\Big]{-\log p(x^n) + \alpha L_n(p)}.
\end{align}
This satisfies Kraft's inequality on $\Xset^n$.

When $x^n \in \Xset^n$ is given,
let $\ddot{p}$ $=\ddot{p}_{x^n}$
denote the minimizer of \eqref{definition_L2p}, i.e.\
\begin{align}
  \label{definition_MDL_estimator}
  \ddot{p} = \ddot{p}_{x^n} = \argmin_{p \in \ddotPset_n} \pair[\Big]{-\log p(x^n) + \alpha L_n(p)}.
\end{align}
Then, we can interpret $\Ltp$ as the codelength of a two part code, i.e.\
we encode $\ddot{p}$ with the code length $\alpha L_n(\ddot{p})$,
then
we encode $x^n$ by the Shannon code corresponding to $\ddot{p}$. 
Let $\ptp =p_{\rm twopart}$ 
denote the sub probability distribution corresponding to $\Ltp$ as
\begin{align}
  \ptp(x^n) = e^{-\Ltp(x^n)} = \ddot{p}(x^n) e^{- \alpha L_n(\ddot{p})}.
\end{align}

The code length function $\Ltp$ depends on 
$(\ddotPset_n, \alpha L_n)$ for each $n$.
Hence in this paper,
we call a code with code length function $\Ltp$ as ``two part code $\Ltp$ based on $(\ddotPset_n, \alpha L_n)$''.
We call 
the function which maps $x^n$ to $\ddot{p}_{x^n}$
as the MDL estimator defined by the two part code $\Ltp$ based on $(\ddotPset_n, \alpha L_n)$.

It is known that
a two part code with small redundancy gives a good MDL estimator.
To state it, here we give the definition of  
R\'enyi divergence and reduncancy.
R\'enyi divergence
of
degree $\lambda \in (0, 1)$, which is a measure of differences between two probability densities,
between $p$ and $q$  defined as
\begin{align}
  \label{definition_Renyi_divergence}
  \dbar_\lambda (p \| q) = - \frac{1}{1 - \lambda} \log \E_p \pair[\Big]{\frac{q(X)}{p(X)}}^{1 - \lambda}.
\end{align}
The R\'enyi divergence is an increasing and continuous function of $\lambda$,
and the limit $\lim_{\lambda \to 1} \dbar_\lambda(p \| q)$ equals the Kullback-Leibler divergence below.
\begin{align}
  \label{definition_KL_divergence}
  D(p \| q) = \E_p \log \frac{p(X)}{q(X)}.
\end{align}
Define the redundancy of a Shannon code corresponding to $p$ 
with respect to $p^*$ as
\begin{align}
  \RED^{(n)} (p^*, p) = \E_{X^n \sim p^*} \interval[\Big]{\log \frac{p^*(X^n)}{p(X^n)}}.
\end{align}
If 
\[
\sup_{p^* \in \Mset} \RED^{(n)}(p^* , \ptp)
\]
is finite for each $n$, let
$\overline{\RED}(n)$ denote it.

The following holds.
\begin{thm}[Barron and Cover 1991, Li 1999, Zhang 2004, Gr\"unwald 2007]
  \label{BC_thm1}
  Let $p^*$ be an arbitrary probability density over $\Xset$.
  For arbitrary $\alpha > 1, \lambda \in (0, 1 - \alpha^{-1}], \ddotPset_n$ and $L_n$,
  the MDL estimator defined by $(\ddotPset_n, \alpha L_n)$ satisfies the following inequalities.
  \begin{align}
    \label{ineq_BC_thm1}
    \E_{X^n \sim p^*} \dbar_\lambda(p^* \| \ddot{p}_{X^n}) \leq \frac{1}{n} \RED^{(n)} (p^*, \ptp)
    \le 
    \min_{\bar{p}\in \ddotPset_n}\Bigl( D(p^*\| \bar{p}) + \frac{1}{n}L_n(\bar{p}) \Bigr).
  \end{align}
\end{thm}

\

We provide a proof in Appendix~\ref{app1}.

The last side of \eqref{ineq_BC_thm1} is the index of resolvability.
If
\begin{align}
  \lim_{n \to \infty} \frac{\overline{\RED}(n)}{n} = 0,
\end{align}
then 
Theorem~\ref{BC_thm1}
implies that
the risk of the MDL estimator defined by this two part code converges to $0$,
uniformly for $p^* \in \Mset$.
This kind of result on MDL estimators first appeared in \cite{Barron_Cover},
where the bound in \cite{Barron_Cover} is given as the index of resolvability.
That theorem and its proof have been improved in \cite{Li_estimation_mixture} and \cite{Zhang_convergence_MDL}.
Theorem~\ref{BC_thm1} is of the form in \cite{Grunwald}.
The book \cite{Grunwald} also provides a historical note on this theorem.

A probabilistic performance guarantee is also known \cite{Zhang_2006,Chatterjee_Barron}.
\begin{thm}[Zhang 2006, Chatterjee and Barron 2014]
  \label{BC_thm2}
  Let $p^*$ be an arbitrary distribution on $\Xset$.
  For arbitrary $\alpha > 1, \lambda \in (0, 1 - \alpha^{-1}], b > 0, \ddotPset_n$ and $L_n$,
  the MDL estimator defined by $(\ddotPset_n, \alpha L_n)$ satisfies the following inequality.
  \begin{align}
    \label{ineq_BC_thm2}
    \Pr \set[\Big]{\dbar_\lambda (p^* \| \ddot{p}_{X^n}) > \frac{1}{n} \log \frac{p^*(X^n)}{\ptp(X^n)} + b} < e^{-n \alpha^{-1} b},
  \end{align}
  where $\Pr$ is measured with $X^n \sim p^*$.
\end{thm}
This theorem provides an upper bound,
which is exponentially small in $n$,
on the probability of the event that
the loss exceeds the value
\begin{align}
  - \log \ptp(x^n) - (- \log p^*(x^n))
\end{align}
by $b$. 

We provide a proof in Appendix~\ref{app2}.

Theorems
\ref{BC_thm1}
and
\ref{BC_thm2}
 imply that
a two part code which enjoys a small regret provides an MDL estimator with small 
loss (with a high probability) and risk,
because a code with small regret shows a small redundancy as well.
Hence we pursue a two-stage code with small regret
for non-exponential families.

The regret of a code $p$ with respect to $\Mset$ and $x^n$ is defined by
\begin{align}
  \label{definition_regret}
  \REG(p; \Mset, x^n) = \max_{q \in \Mset} \set[\Big]{-\log p(x^n) - \pair[\Big]{- \log q(x^n)}}.
\end{align}
Then, since the maximum over $q \in \Mset$ is greater than at $p^* \in \Mset$,
\begin{align}
  \RED^{(n)}(p^*, p) \leq \E_{p^*} \REG(p; \Mset, x^n)
\end{align}
holds.

In this paper, we assume that $\Mset$ is a signle parametric model
$
  \Mset = \set{p_\theta : \theta \in \Theta \subseteq \Rset^K} \hspace{1mm} (K \ge 1).
$
Gr\"unwald \cite{Grunwald} provides a way of constructing a two part code
which achieves the regret bound \eqref{intro_Grunwald_regret}. 
It is close to the minimax value \eqref{asymptotic_minimax_regret}, for exponential families.
In the next section,
we will introduce the way of this construction and show its properties 
which lead to a nice regret bound.
Further, in the section \ref{section_extension_to_non_exponential},
we will develop an extension of this method so that
we can obtain a small upper bound on the regret for non exponential families.

\section{Gr\"unwald's construction of two part codes for exponential families}
In this section,
we construct
a two part code $(\ddot{\Theta}_n, L_n)$ which is appropriate for our goal,
following Gr\"unwald \cite{Grunwald}.
Then, we give an analysis on the code length of the two part code.
This analysis is slightly finer than that of \cite{Grunwald} and gives
a bound for finite sample size.
For the exponential families case,
we show that the two part code has an upper bound of the regret
close to the minimax regret.

In this paper, we assume
$\Mset$ is a $K$ dimensional parametric model
\begin{align}
  \Mset = \set{p_\theta : \theta \in \Theta},
\end{align}
where
$\Theta$ is a convex and compact subset of $\Rset^K$.
We denote the maximum likelihood estimate with respect to $\Theta$ as
\begin{align}
  \label{definition_MLE}
  \thetahat = \thetahat(x^n) = \argmax_{\theta \in \Theta} p_\theta(x^n).
\end{align}
In this setting,
we may assume that 
$\Theta$ is a subset of a naturally determined parameter space $\tilde{\Theta}$.
Here, ``naturally determined'' means that the space cannot be extended any more.
For example, the probability parameter for the Bernoulli model
has the natural range $[0,1]$
and $\Theta$ may be an interval contained in $[0,1]$.
Then, when $\thetahat$ is on the boundary of $\Theta$,
$\thetahat$ is not always a stationary point of the likelihood function
over $\tilde{\Theta}$.
In that case, the essential maximum likelihood estimate, which is with respect to $\tilde{\Theta}$,
is in $\tilde{\Theta} \setminus \Theta$.
However, in this paper, we always let the maximum likelihood estimate refer to \eqref{definition_MLE}
which is with respect to the restriction $\Theta$.

First, we put some assumptions on the model $\Mset$
for our analysis.
We assume the differentiability of the likelihood function.
\begin{assume}
  \label{assumption_cont_differentiability}
  For all $x^n$ and for all $\theta \in \Theta$, $\log p_\theta(x)$ is $3$ times continuously differentiable with respect to $\theta$.
\end{assume}
Define the empirical Fisher information matrix by
\begin{align}
  \Jhat_{ij}(\theta; x^n) = - \frac{1}{n} \sum_{t = 1}^{n} \frac{\partial^2}{\partial \theta_i \partial \theta_j} \log p_{\theta}(x_t)
\end{align}
and let $\Jhat(\theta; x)$ denote it for the case of $n=1$.
Further, we define the Fisher information matrix $J(\theta)$ as the expectation of $\Jhat(\theta; x)$ as $J(\theta) = \E_{p_\theta} \Jhat(\theta; X)$.
Then, we assume that for all $\theta \in \Theta$, $J(\theta)$ is positive definite and the eigenvalues of $J(\theta)$ 
are bounded on $\Theta$.
\begin{assume}
  \label{assumption_J_eigen_values}
  There exist positive $\zeta$ and $\lambdabar$ such that for all unit vectors $z$ and for all $\theta \in \Theta$, the following holds.
  \begin{align}
    \label{ineq_assumption_J_eigen_values}
    \zeta \leq z^T J(\theta) z \leq \lambdabar.
  \end{align}
\end{assume}
\begin{assume}
  \label{assumption_J_determinant}
  The determinant $\abs{J(\theta)}$ is a differentiable function on $\Theta$.
\end{assume}
\begin{assume}
  \label{assumption_D_J}
  There exists $D_J > 0$ such that
  \begin{align}
    \sup_{\abs{z} = 1, \theta \in \Theta} \abs[\Big]{\Del_z \abs{J(\theta)}^{1/2}} \leq D_J,
  \end{align}
  where $\Del_z$ means the directional derivative with respect to 
  $\theta$ to the direction of $z$.
\end{assume}
{
\begin{assume}
    \label{assumption_kappa}
There exist $\kappa > 0$ and $\bar{b} >0$ such that
for all $(\theta_1,\theta_2) \in \Theta^2$ 
with $|\theta_1 - \theta_2| \le \bar{b}$, the following holds.
\[
\max_{z \in \Rset^K\setminus \{ 0 \}}\frac{z^TJ(\theta_1)z}{z^TJ(\theta_2)z} \leq 1+\kappa |\theta_1 - \theta_2|.
\]
\end{assume}
}

To handle the sequences $x^n$ for which the maximum likelihood estimate 
is on the boundary of $\Theta$, denoted by $\boundary{\Theta}$,
we use the following assumption.
\begin{assume}
  \label{assumption_boundary}
  It is possible to construct a 
  two part code $\Ltp^\partial(x^n) = -\log \ptp^{\partial}(x^n)$ 
  over $\{x^n : \hat{\theta} \in \partial \Theta \}$
  based on $(\ddotPset^\partial_n, \alpha L^{\partial}_n)$,
   such that,
  for a certain $d > 0$, for all large $n$,
  for all $\alpha > 0$, for all $x^n$ with $\hat{\theta} \in \partial \Theta$,
  \[
  \Ltp^\partial(x^n)-\log \frac{1}{p_{\hat{\theta}}(x^n)}
  \le  \frac{\alpha(K-d)}{2}\log n + o(\log n)
  \]  
  holds.
\end{assume}

{\it Remark:}
If $\partial \Theta$ consists of hyper surfaces in $\Rset^K$, 
for example
a surface of a polyhedron or that of a sphere, 
it is easy to construct such $\ptp^{\partial}$ with $d=1$.

The code we investigate uses a quantization of $\Theta$ based on the Fisher 
information matrix.
This quantization, which we quote from Gr\"unwald (Chapter 10 of \cite{Grunwald}),
originates with the analysis by Barron \cite{barron1985logically}.

Let $\ddot{\Theta}_n \subseteq \Theta$ 
be a quantized version of $\Theta$
and
$\ddotPset_n = \set{p_\theta : \theta \in \ddot{\Theta}_n}$.
Further, let $L_n$ be a code length function on $\ddot{\Theta}_n$. 
When $\Mset$ is an exponential family of densities,
we can show a small upper bound on the regret of the two part code.

Fix $a > 0$.
Here we introduce the construction of the two part code in \cite{Grunwald}.
First, we partition each axis of $\Rset^K$ into
$\{ [(i - 1) a n^{-\beta}, i a n^{-\beta})\}_{i \in \Zset}$ with $0<\beta<1/2$.
This partition defines hyper cubes with side lengths $a n^{-\beta}$.
Note that Gr\"unwald assumes $\beta = 1/4$, which is turned to provide the best bound later in our analysis. 

We construct a cover of $\Theta$ which consists of these hyper cubes.
We call the intersections of $\Theta$ and these hyper cubes large cells.
For each $x^n$, we define $\theta_S = \theta_S(x^n)$ as the center of mass of the large cell
which includes $\thetahat(x^n)$.
Then, for all $x^n$, if $\theta$ belongs to
the large cell including $\theta_S(x^n)$, then
\begin{align}
  \abs{\theta - \theta_S(x^n)} \leq \frac{\sqrt{K}a}{2} n^{-1/4}
\end{align}
holds.
For each large cell $\Scell$,
let $\theta_{\Scell}$ be the center of mass of $\Scell$.
Since the Fisher information matrix $J(\theta_{\Scell})$ is positive definite by Assumption~\ref{assumption_J_eigen_values},
we can define the ellipsoid 
\begin{align}
  \label{definition_ellipsoid}
  \set[\Big]{\theta : (\theta - \theta_{\Scell})^T J(\theta_{\Scell}) (\theta - \theta_{\Scell}) \leq \frac{a^2}{4n}}
\end{align}
with the center $\theta_{\Scell}$.

For each large cell $\Scell$, consider the hyper rectangle which circumscribes
the ellipsoid \eqref{definition_ellipsoid}
and whose edges are parallel to it there axes.
This is a small hyper rectangle centered in $\Scell$.
Next we construct a cover of each $\Scell$ using
disjoint copies of that hyper rectangle with center shifted to cover $\Scell$.


We call the intersection of $\Theta$ and each of the hyper rectangles a small cell.
Then for each small cell, we put a quantized point on 
its center of mass.
Let $\ddot{\Theta}_n$ denote the set of all such quantized points.
Then, define $L_n$ as the code length of the fixed length code for $\ddot{\Theta}_n$:
\begin{align}
  L_n(\theta) = \log \abs{\ddot{\Theta}_n}.
\end{align}

Now, define the MDL estimator induced by our $(\ddot{\Theta}_n,\alpha L_n)$ as
\begin{align}
  \ddot{\theta} = \ddot{\theta}(x^n) = \argmin_{\theta \in \ddot{\Theta}_n} \pair[\Big]{- \log p_\theta(x^n) + \alpha L_n(\theta)},
\end{align}
which is the point corresponding to the MDL estimate $\ddot{p}$.
Then, the code length of the two part code is
\begin{align}
  - \log \ptp(x^n) = - \log p_{\thetaddot}(x^n) + \alpha L_n(\thetaddot).
\end{align}

Under Assumptions~\ref{assumption_cont_differentiability} to \ref{assumption_kappa},
we can show the following proposition and lemma
\begin{proposition}
  \label{prop_quantization}
  Suppose Assumptions~\ref{assumption_cont_differentiability} to \ref{assumption_kappa} hold.
  Let $\thetaddh$ denote any of the closest points in  
  $\ddot{\Theta}_n$
  to $\thetahat$.
  The quantization $\ddot{\Theta}_n$ satisfies the following inequalities
  for all $x^n$.
  \begin{align}
    (\thetaddh - \thetahat)^T J(\theta_S) (\thetaddh - \thetahat) \leq \frac{Ka^2}{4n}.
    \label{quantize_ineq1}
  \end{align}
  Further, for an arbitrary point on the line $\theta$ between $\thetaddh$ and $\thetahat$, we have
  \begin{align}
    \abs{\theta - \theta_S} \leq \sqrt{K} a n^{-1/4}.
    \label{quantize_ineq2}
  \end{align}
\end{proposition}
\begin{lemma}
  \label{prop_L_n}
    Suppose Assumptions~\ref{assumption_cont_differentiability} to \ref{assumption_kappa} hold.
  The code length function $L_n(\theta) = \log \abs{\ddot{\Theta}_n}$ on $\ddot{\Theta}_n$ satisfies
  the following inequality for all 
  { $n \geq (\sqrt{K} a / \bar{b})^{1/\beta}$}
  and
  for all $\theta \in \ddot{\Theta}_n$,
  \begin{align}
    L_n(\theta)
    \leq \frac{K}{2} \log n + \log \int_{\Theta} \abs{J(\theta')}^{1/2} d\theta' - K\log a
    + r(n),
    \label{parameter_length}
  \end{align}
  where {
  \begin{align}\label{def_of_r}
    r_\beta(n) =
    \log (1 \!+\! C_J n^{-(1/2-\beta)}) + \log (1 \!+\! C_{\Theta} n^{-\beta}) + \log (1 \!+\! C_{J, K} n^{-\beta}),
  \end{align}}
  and $C_J, C_{\Theta}$ and $C_{J, K}$ are the constants provided in the proof.
\end{lemma}
The proofs of the proposition and the lemma stated above are given in Section~\ref{section_proofs}.
Note that $r_\beta(n)$ is decreasing and $\lim_{n \to \infty} r_\beta(n) = 0$.
We see in this expression for $r(n)$ that $\beta = 1/4$
provides the best bound with $r(n)=O(n^{-1/4})$.

The following lemmas describe properties of the quantized space $\ddot{\Theta}_n$.
\begin{lemma}
  \label{lemma_quad_J_thetahat_bound}
  Let $\thetaddh$ denote any of the closest points in  
  $\ddot{\Theta}_n$
  to $\thetahat$.
  Under Assumption~\ref{assumption_kappa},
  for all 
  {$n \geq (\sqrt{K} a / \bar{b})^{1/\beta}$}
  ,
  \begin{align}
  \sup_{x^n : \thetahat \in \Theta^\circ}
    (\thetaddh - \thetahat)^T J(\thetahat) (\thetaddh - \thetahat)
    \leq
    \frac{Ka^2}{4n} \pair{1 + \kappa \sqrt{K} a n^{-\beta}}.
    \label{quantize_ineq3}
  \end{align}
\end{lemma}

\begin{IEEEproof}[Proof of Lemma~\ref{lemma_quad_J_thetahat_bound}]
  Since $\thetahat$ and $\theta_S$ belong to the same large cell with side length $a n^{-\beta}$,
  \begin{align}
    \abs{\thetahat - \theta_S} \leq \sqrt{K} a n^{-\beta}
  \end{align}
  holds.
  Therefore, 
  from Assumption~\ref{assumption_kappa},
  when $n \geq (\sqrt{K} a / \bar{b})^{1/\beta}$,
  we have for all $\theta$ which belongs to the same large cell as $\thetahat$,
  \begin{align}
    (\theta - \thetahat)^T J(\thetahat) (\theta - \thetahat) 
    \leq
    (\theta - \thetahat)^T J(\theta_S) (\theta - \thetahat) \pair[\Big]{1 + \kappa \sqrt{K} a n^{-\beta}}.
    \label{Taylor_expansion_bound_J}
  \end{align}
  From the fact that $\thetahat$ and $\thetaddh$ belong to the same small cell defined by using $J(\theta_S)$,
  \begin{align}
    \label{quantize_ineq_theta_S}
    (\thetaddh - \thetahat)^T J(\theta_S) (\thetaddh - \thetahat) \leq \frac{Ka^2}{4n}
  \end{align}
  holds.
  By plugging in $\thetaddh$ for $\theta$ in \eqref{Taylor_expansion_bound_J}, we have \eqref{quantize_ineq3}.
\end{IEEEproof}

\begin{lemma}
  \label{lemma_data_description_loss_bound}
  Under Assumptions~\ref{assumption_D_J} and \ref{assumption_kappa}, the following inequality holds
  for all 
  {$n \geq (\sqrt{K} a / \bar{b})^{1/\beta}$}, 
  for all $x^n$, and for all $\theta'$ between $\thetaddh$ and $\thetahat$.
  \begin{align}
    (\thetaddh - \thetahat)^T J(\theta') (\thetaddh - \thetahat)
    \leq
    \frac{C_n Ka^2}{4n},
    \label{ineq_data_description_loss_bound}
  \end{align}
  where
  \begin{align}
    C_n = \pair{1 + \kappa \sqrt{K} an^{-\beta}} \pair{1 + \kappa \sqrt{K} a n^{-1/2} \zeta^{-1/2} / 2}.
    \label{definition_C_n}
  \end{align}
\end{lemma}

\begin{IEEEproof}[Proof of Lemma~\ref{lemma_data_description_loss_bound}]
  By using Assumption~\ref{assumption_kappa} and 
  Lemma~\ref{lemma_quad_J_thetahat_bound}, we have
  \begin{align}
    (\thetaddh - \thetahat)^T J(\theta') (\thetaddh - \thetahat)
    \leq
    \frac{Ka^2}{4n} (1 + \kappa \sqrt{K} an^{-\beta})(1 + \kappa \abs{\theta' - \thetahat}).
  \end{align}
  Since $\theta'$ and $\thetahat$ belong to the same small cell,
  from Assumption~\ref{assumption_J_eigen_values},
  $\abs{\theta' - \thetahat}^2 \leq Ka^2/4n \zeta$ holds.
  Then,
  \begin{align}
    (\thetaddh - \thetahat)^T J(\theta') (\thetaddh - \thetahat) 
    \leq
    \frac{Ka^2}{4n} (1 + \kappa \sqrt{K} an^{-\beta}) (1 + \kappa \sqrt{K} an^{-1/2} \zeta^{-1/2} / 2).
  \end{align}
\end{IEEEproof}

Since $\thetaddh$ may not be the minimum solution to the description length,
the regret of the two part code is bounded by
\begin{align}
  \REG(\ptp, \Mset, x^n) \leq \log \frac{p_{\thetahat}(x^n)}{p_{\thetaddh}(x^n)} + \alpha L_n(\thetaddh).
\end{align}
By the Taylor's theorem, for $x^n$ with $\thetahat \in \interior{\Theta}$, there exists $\theta'$ between $\thetaddh$ and $\thetahat$
such that
\begin{align}
  \log \frac{p_{\thetahat}(x^n)}{p_{\thetaddh}(x^n)} =
  \frac{n}{2} (\thetaddh - \thetahat)^T \Jhat(\theta'; x^n) (\thetaddh - \thetahat).
  \label{regret_bound_by_Jhat_pre}
\end{align}
Then, we have
\begin{align}
  \REG(\ptp, \Mset, x^n) \leq 
  \frac{n}{2} (\thetaddh - \thetahat)^T \Jhat(\theta'; x^n) (\thetaddh - \thetahat) + \alpha L_n(\thetaddh).
  \label{regret_bound_by_Jhat}
\end{align}
When $\Mset$ is an exponential family and $\theta$ is its canonical parameters,
$\Jhat(\theta; x^n) = J(\theta)$ holds for all $\theta \in \Theta$ and for all $x^n$.
Using this fact, we can show
the following lemma from Lemma~\ref{prop_L_n} and Lemma~\ref{lemma_data_description_loss_bound}.
\begin{lemma}
  \label{lemma_exponential_family_regret_bound}
  When $\Mset$ is an exponential family and $\Theta$ is the range of its canonical parameters,
  our two part code has the regret which satisfies the following.
  For all $x^n$ with $\thetahat \in \interior{\Theta}$,
  \begin{align}
    \label{regret_exponential_family}
    \frac{1}{\alpha} \REG(\ptp; \Mset, x^n) 
    \leq
    \frac{K}{2} \log n + \log \int_{\Theta} \abs{J(\theta)}^{1/2} d\theta
    + f(n),
  \end{align}
  where
  \begin{align}
    f(n)
    =  \frac{C_n Ka^2}{8\alpha}- K \log a 
    + r(n),
  \end{align}
  $r(n) (=O(n^{-1/4}))$ denotes 
  $r_{\beta}(n)$ 
  \eqref{def_of_r} with $\beta=1/4$, and $C_n (= 1 + o(1))$ is defined as \eqref{definition_C_n}.
\end{lemma}
The proofs of Proposition~\ref{prop_quantization} and Lemma~\ref{prop_L_n} and Lemmas~\ref{lemma_quad_J_thetahat_bound} and
\ref{lemma_data_description_loss_bound} will be provided in Section~\ref{section_proofs}.

\subsection{Note on boundary problem}
The code $\ptp$ is good for $x^n$ with $\hat{\theta} \in \interior{\Theta}$,
but \eqref{regret_exponential_family} is not guaranteed for $x^n$
with $\thetahat \not\in \interior{\Theta}$.
That is, $\ptp$ is not enough for our purpose,
and we need a bound for all $x^n$.
For this point, 
important is Assumption~\ref{assumption_boundary},
which requires existence of 
a code $\ptp^{\partial}$ with small regret for $x^n$ with $\hat{\theta} \in {\partial\Theta}$.
That is, under Assumption~\ref{assumption_boundary},
we can modify $\ptp$  
with the help from $\ptp^{\partial}$,
so that the regret bound \eqref{regret_exponential_family} holds also when $\thetahat \in \boundary{\Theta}$.

Let $l_1(n)=n^{-\iota}$ with $d > 2 \iota > 0$.
Define
\begin{align}
  - \log \ptp'(x^n) =
  \begin{cases}
    - \log \ptp(x^n) + \alpha l_1(n), & \text{if $\thetahat \in \interior{\Theta}$}, \\
    -\log \ptp^\partial(x^n)
    - \alpha \log (1 - e^{-l_1(n)}), & \text{if $\thetahat \in \boundary{\Theta}$}.
  \end{cases}
  \label{definition_combined_two_part_code}
\end{align}
In this scheme, 
$\ptp'(x^n)$
employs $\ptp$ when $\thetahat \in \interior{\Theta}$, otherwise it uses $\ptp^{\partial}$.
The terms $\alpha l_1(n)$
and $- \alpha \log (1 - e^{-l_1(n)})$
are extra code length to indicate 
whether we use $\ptp'$ or $\ptp^{\partial}$.
We employ the MDL estimator defined by $\ptp'$ for our purpose. 
Note that that MDL estimator outputs the same estimate as that by $\ptp$, when 
$\hat{\theta} \in \Theta^\circ$,
and the same one as by $\ptp^{\partial}$, otherwise.

Now assume that $n$ satisfies
$l_1(n) = n^{-\iota}  \le 1/2$ (i.e.\ $n \ge 2^{1/\iota}$), then
$e^{-l_1(n)} \le 1-l_1(n)/2$ holds.
Then, when $n \ge 2^{1/\iota}$, the following holds.
\[
-\log (1 - e^{-l_1(n)}) \le -\log (1 - 1 + l_1(n)/2) = \iota \log n +\log 2,
\]
which implies that,
if $\hat{\theta} \in \partial \Theta$,
\[
   \log \frac{p_{\hat{\theta}}(x^n)}{\ptp'(x^n)}=\log \frac{p_{\hat{\theta}}(x^n)}{\ptp^\partial(x^n)}
   +\alpha \log (1-e^{-l_1(n)})
   \le \alpha \Bigl( \frac{K-(d - 2 \iota )}{2}\log n +  \log 2\Bigr),
\]
holds.
As for the case that $\hat{\theta} \in \Theta^\circ$, we have
\[
\log \frac{p_{\hat{\theta}}(x^n)}{\ptp'(x^n)}
\le 
\alpha \Bigl(\frac{K}{2}\log n 
+  \log \int_\Theta |J(\theta)|^{1/2} +  f(n) +  l_1(n)\Bigr).
\]
Here, the former is smaller than the latter, when $n$ is large. (Recall $2\iota < d$ is assumed.)
Let
\begin{align}
  \overline{\REG}
  =
  \alpha \pair[\Big]{
  \frac{K}{2} \log n + \log \int_{\Theta} \abs{J(\theta)}^{1/2} d\theta
  + f(n) + l_1(n)
  }.
  \label{REG_for_exponential_families}
\end{align}
Then for large $n$, $\REG(\ptp', \Mset, x^n) \leq \overline{\REG}$ holds for all $x^n$.
Therefore, by Theorems~\ref{BC_thm1} and \ref{BC_thm2}, we have the following.
\begin{cor}[Bounds for exponential families]
  \label{BC_cor}
  Let $\Mset$ be an exponential family which satisfies Assumptions~\ref{assumption_cont_differentiability}--\ref{assumption_boundary}.
  Let $\ddot{p}_{X^n}$ be the estimate by
  the MDL estimator induced by $\ptp'$.
  When $X^n$ is drawn from $p^* \in \Mset$, for large $n$, we have
  \begin{align}
    \label{ineq_BC_cor1}
    \E_{X^n \sim p^*} \dbar_\lambda(p^* \| \ddot{p}_{X^n}) \leq \frac{1}{n} \overline{\REG}
  \end{align}
  and 
  \begin{align}
    \label{ineq_BC_cor2}
    \Pr \set[\Big]{\dbar_\lambda (p^* \| \ddot{p}_{X^n}) > \frac{1}{n} \overline{\REG} + b} < e^{-n \alpha^{-1} b},
  \end{align}
  where $b$ is an arbitrary positive real and 
  $\Pr$ is defined for $X^n \sim p^*$.
\end{cor}

{\it Remark:} 
The analysis may be generalized to two part codes with possible non-compact
parameter spaces taking advantage of a continuous prior probability density function $w(\theta)$.
The parameter space may be split into moderately small regions such that
within each region, by continuity,
$
w(\theta)
$
and
$|J(\theta)|$
are nearly constant
(these regions being described with length the minus log of their prior probability).
These regions are further quantized into the small cells discussed above
(described with length the log cardinality of them in each region).
Then the regret
$
\Ltp(x^n)-\log (1/p_{\hat{\theta}}(x^n))
$
in an exponential family has the bound
\[
\frac{K}{2}\log n + \log \frac{|\hat{J}(\hat{\theta})|^{1/2}}{w(\hat{\theta})}+\frac{Ka^2}{8}-K\log a +O\Bigl(\frac{1}{n}\Bigr).
\]
In the case that $\int |J(\theta)|^{1/2}d\theta$ is finite
the conclusion of Lemma 4 arises with the choice of Jeffreys prior
\[
w(\theta)=\frac{|J(\theta)|^{1/2}}{\int |J(\theta)|^{1/2}d\theta}.
\]

\section{Extension to Non-exponential families}
\label{section_extension_to_non_exponential}

We have given the upper bound \eqref{REG_for_exponential_families} of the regret  for exponential families  which is close to optimal.
What about non-exponential families?
We find that for non-exponential families,
one cannot obtain
similar upper bound by the same two part code as for exponential-families.
However,
we can improve this two part code using a fiber bundle of local exponential families \cite{Amari_Nagaoka}
so that we can obtain a regret bound similar to \eqref{intro_Grunwald_regret} under some assumptions. 

The inequality \eqref{regret_bound_by_Jhat} itself still holds
 for non-exponential families,
but 
$\Jhat(\theta; x^n)$ differs from $J(\theta)$ in general for any parameterization.
In particular it does even at $\theta = \hat{\theta}$.
Therefore, we have to deal with the difference between them to evaluate an upper bound on $(\thetaddh - \thetahat)^T \Jhat(\theta; x^n) (\thetaddh - \thetahat)$ for non-exponential families.
In this section we describe our recipe based on fiber bundle of local exponential families
for the problem.

\subsection{Fiber Bundle of Local Exponential Families}

Following \cite{Takeuchi_Barron_2013}, we define a random variable $V(\theta; x^n)$ 
 which represents the difference between Fisher information and empirical 
 Fisher information:
\begin{align}
  V(\theta; x^n) = J^{-1/2}(\theta)^T \Jhat(\theta; x^n) J^{-1/2}(\theta) - I,
\end{align}
where $I$ is the unit matrix of order $K$.
Note that $\E_{p_\theta}  V(\theta; x^n) =J^{-1/2}JJ^{-1/2}-I =0$.
Let $\Xi_0 = [-\xi_0, \xi_0]^{K \times K}$ for a constant $\xi_0 > 0$.
For each $\theta \in \Theta$ and each $\xi \in \Xi_0$,
we define the following density:
\begin{align}
  \label{definition_local_exponential_family_bundle}
  \pbar_{\theta, \xi}(x) &=  p_\theta(x) \exp \pair[\Big]{\xi \cdot V(\theta; x) - \psi_{\theta}(\xi)},\\
  \psi_{\theta}(\xi) &= \log \int p_{\theta}(x) \exp \pair[\Big]{\xi \cdot V(\theta; x)} dx.
\end{align}
Here, $\xi \cdot V(\theta; x)$ denotes the
Frobenius inner product of $\xi$ and $V(\theta; x)$.
Let 
$\pbar_{\theta, \xi}(x^n) 
=
\prod_{t=1}^n\pbar_{\theta, \xi}(x_t)$.
Note that 
\begin{align}
\pbar_{\theta, \xi}(x^n) 
=  p_\theta(x^n) \exp 
\pair[\Big]{n \big(\xi \cdot V(\theta; x^n) - \psi_{\theta}(\xi)\big)}
\label{bundlecode}
\end{align}
holds.
For each $\theta$, the family $\Mset_e(\theta) = \set{\pbar_{\theta, \xi} : \xi \in \Xi_0}$ is an exponential family with the natural parameter $\xi$.
When $\xi = 0$, $\pbar_{\theta, \xi}$ equals $p_{\theta}$.
The structure where for each $\theta$, the exponential family $\Mset_e(\theta)$ is supplemented to $\Mset$ forms a fiber bundle.
This fiber bundle is called as a fiber bundle of local exponential families \cite{Amari_Nagaoka}.
We use $\set{\pbar_{\theta, \xi} : \theta \in \Theta, \xi \in \Xi_0}$ to construct our two part code.

\subsection{Two-Part Codes based on Fiber Bundle of Local Exponential Families}

We consider a finite subset $\Xi_n \subset \Xi_0$, a quantized version of $\Xi_0$.
Let $\tilde{L}_n(\xi)$ be a code length function on $\Xi_n$ satisfying Kraft's inequality.
Let $\ddot{\Theta}_n$ and $L_n$ be the same one as what we constructed in the previous section.
We employ a parameter space $\ddot{\Theta}_n \times \Xi_n$ for the two part code using $\set{\pbar_{\theta, \xi}}$.
Let $\Lbar_n(\theta, \xi) = L_n(\theta) + \tilde{L}_n(\xi)$.
The two part code is denoted as $(\ddot{\Theta}_n \times \Xi_n, \alpha \Lbar_n)$.

Given the two part code,
we define an MDL estimator as
\begin{align}
  (\ddot{\theta}, \ddot{\xi})
  = \argmin_{(\theta, \xi) \in \ddot{\Theta}_n \times \Xi_n} \pair[\Big]{-\log \pbar_{\theta, \xi}(x^n) + \alpha \Lbar_n(\theta, \xi)}.
\end{align}
Note that the estimate 
$\pbar_{\ddot{\theta}, \ddot{\xi}}$
may be
 a distribution outside $\Mset$.
Define a sub probability distribution corresponding to the modified two part code as
\begin{align}
  \ptpbar(x^n) = \pbar_{\thetaddot, \xiddot}(x^n) e^{-\alpha \Lbar_n(\thetaddot, \xiddot)}.
\end{align}

We construct $\Xi_n$ so that the new two part code has a good upper bound of its regret.
Define the matrix $\mathcal{E}^{(l,m)}$ for each $(l,m)$ as
\begin{align}
  \interval{\mathcal{E}^{(l,m)}}_{ij} = \Indi((i, j) = (l, m)),
\end{align}
where $\Indi(\cdot)$ is the indicator function.
Then, using a positive decreasing sequence $\set{u_n}$ (we determine later), let
\begin{align}
  \Xi_n = \set{0} \cup \set{ \pm u_n \mathcal{E}^{(l,m)} : l, m \in \set{1,2,\cdots, K}}.
\end{align}
Here, $\Xi_n$ for each $n$ consists of matrices with single non-zero value $u_n$ or $-u_n$ and the zero matrix.
Therefore, $\abs{\Xi_n} = 2K^2 + 1$.
We define the maximum norm of matrices defined as
\begin{align}
  \dabs{V}_M = \max_{i, j} \abs{V_{ij}}.
\end{align}
Note that for all $\theta \in \Theta$ and for all $x^n$,
we can select $\xi \in \Xi_n$ so that
\begin{align}
  \xi \cdot V(\theta; x^n) = u_n \dabs{V(\theta; x^n)}_M.
  \label{selection_xi}
\end{align}
Actually,
when the $(i, j)$ element of $V(\theta; x^n)$ satisfies
$\abs{V_{ij}(\theta; x^n)} = \dabs{V(\theta; x^n)}_M$,
either of $\pm u_n \mathcal{E}^{(i, j)} \in \Xi_n$
with the same signature as the element achieves \eqref{selection_xi}.
Note that
by definition of the maximum norm,
such index $(i, j)$ always exists for $V(\theta; x^n) \neq 0$.

Since $\abs{\Xi_n} = 2K^2 + 1$, to encode $\xi \in \Xi_n$,
we could use the uniform code length $\log (2K^2 + 1)$.
However, we employ a variable length code to obtain a smaller regret bound.
Let $l_2(n) = n^{-\nu}$ ($\nu > 0$) 
and
\begin{align}
  \bar{l}_2(n) = \log \frac{1}{1 - e^{-l_2(n)}} + \log (2K^2).
\end{align}
We design
\begin{align}
  \tilde{L}_n(\xi)
  =
  \begin{cases}
    l_2(n), & \text{if $\xi = 0$,} \\
    \bar{l}_2(n), & \text{otherwise.}
  \end{cases}
\end{align}
This $L_n(\xi)$ satisfies Kraft's inequality on $\Xi_n$ as the equality.
Now, we have defined a new two part code $(\Theta \times \Xi_n, \alpha \Lbar_n)$
with a hyper parameter $u$.
Note that 
\begin{align}\label{eq:defferenceL2}
    \bar{l}_2(n) \le l_2(n) -\log l_2(n) + \log (2K^2)
\end{align}
holds, which follows from
the inequality $e^{-x} \le 1 - e^{-x}x$ equivalent to
$e^x \ge 1+x$.

We show that
by choosing the value of $u_n$ properly, the two part code has a good regret bound.
The proper value is determined based on some assumptions on $\Mset$.

\subsection{Regret Bounds for the New Two-Part Codes}
To obtain an upper bound of the regret of the new two part code,
we need some preliminaries and assumptions.
Let $\delta_n= (\deltacoef \log n /n)^{1/2}$ ($\deltacoef > 0$).
We define two subsets of $\Xset^n$ as follows:
\begin{align}
  \Gset &= \set{x^n : \dabs{V(\thetahat; x^n)}_M \leq \delta_n, \thetahat \in \interior{\Theta}},\\
  \Gset^c &= \set{x^n : \dabs{V(\thetahat; x^n)}_M > \delta_n, \thetahat \in \interior{\Theta}}.
\end{align}
We need additional assumptions on $\Mset$.
\begin{assume}
  \label{assumption_epsilon_weak}
There exist $\kappa' > 0$ and $\bar{b}'>0$ such that
the following holds,
for all $x^n \in \Gset$,
for all $\theta : |\theta - \hat{\theta}| \le \bar{b}'$, and for all $z \in \Rset^K$.
\[
z^T\hat{J}(\theta,x^n)z 
\leq (1+\kappa' |\theta - \thetahat|)z^T\hat{J}(\hat{\theta},x^n)z.
\]
\end{assume}
This Assumption~\ref{assumption_epsilon_weak}
is used for $x^n \in \Gset$,
while for $x^n \in \Gset^c$, we use the following.
\begin{assume}
  \label{assumption_epsilon}
  There exist $\epsilon > 0$ and $C_{\epsilon} > 0$ such that
  the following holds
  for all $x^n \in \Gset^c$
  and
  for all $z \in \Rset^K$,
  \begin{align}
  \max_{\theta : \abs{\theta -\thetahat}\le \epsilon}
  z^T\Jhat(\theta; x^n)z
  \le
    C_{\epsilon} z^T\pair[\Big]{\Jhat(\thetahat; x^n) + J(\thetahat)} z.
  \end{align}
\end{assume}

\begin{assume}
  \label{assumption_V1}
  There exist certain constants $\Delta > 0$ and $\gamma \in (0,1)$ such that
  for all $x^n \in \Gset^c$,
  \begin{align}
    \inf_{\theta : \abs{\theta - \thetahat} \leq \Delta} \frac{\dabs{V(\theta; x^n)}_M}{\dabs{V(\thetahat; x^n)}_M} > \gamma.
  \end{align}
\end{assume}
Note that
for all $x^n \in \Gset^c$ and for all $\theta$ such that $\abs{\theta - \thetahat} \leq \Delta$, we have
$\dabs{V(\theta; x^n)}_M > \gamma \delta_n$.
\begin{assume}
  \label{assumption_V2}
 There exist  certain constants $\bar{\delta} > 0$ and $B > 0$ such that,
  for all $i,j,k,l$,
  \begin{align}
     \sup_{\theta \in \Theta} 
  \sup_{\xi : \dabs{\xi}_M \leq \bar{\delta}}
    \abs[\big]{\E_{\pbar_{\theta, \xi}} [V_{ij}(\theta; X)V_{kl}(\theta; X)]} \leq B.
  \end{align}
\end{assume}

The following theorem gives an upper bound of 
the code length of our two part code
and
suggests the value of the hyper parameter $u_n$
as $\gamma \delta_n / B$.
\begin{thm}
  \label{thm_improved_regret_bound}
  Under Assumptions~\ref{assumption_cont_differentiability}--\ref{assumption_kappa} and
  \ref{assumption_epsilon}--\ref{assumption_V2},
  the two part code based on $(\ddot{\Theta}_n \times \Xi_n, \alpha \Lbar_n)$
  with $u_n = \gamma \delta_n / B$ and $\gamma g/2B - \nu \alpha > 0$
  has the following regret bound
  for all $n$ satisfying
  \begin{align}\label{condition_n0}
      n > \max \set[\Big]{
        \frac{Ka^2}{4\zeta} \max\set[\Big]{\frac{1}{\epsilon^2}, \frac{1}{\Delta^2}},
\exp\Bigl(\frac{C_\epsilon Ka^2  +\alpha  \log (2K^2)}{\gamma g/2B - \nu \alpha}\Bigr),
        \frac{1}{\kappa^4 K^2 a^4}, \frac{4\zeta}{\kappa^2 K a^2} 
    },
  \end{align}
  \begin{align}
  \sup_{x^n : \hat{\theta} \in \interior{\Theta}}
    \frac{1}{\alpha} \REG(\ptpbar; \Mset, x^n)
    \leq
    \frac{K}{2} \log \frac{n}{2 \pi} + \log \int_{\Theta} \abs{J(\theta)}^{1/2} d\theta
    +
    f_{\rm ne}(n),
    \label{ineq_improved_regret_bound}
  \end{align}
  where
  \begin{align}
    \label{f_delta_c_0}
    f_{\rm ne}(n)
    &=
    \frac{K}{2} \log 2\pi - K \log a +  \frac{C_{\Gset,n} Ka^2}{8 \alpha} (1 + K \delta_n) 
    + r(n) + l_2(n),\\
    C_{\Gset,n}
    &= C_n  (1+\kappa'K^{1/2}a \zeta^{-1/2}n^{-1/2}/2 )
    \label{definition_C_gn},\\
    r(n) &= 
    \log (1 \!+\! C_J n^{-1/4}) + \log (1 \!+\! C_{\Theta} n^{-1/4}) + \log (1 \!+\! C_{J, K} n^{-1/4}).
  \end{align}
\end{thm}

{\it Remark:} 
Note that $\lim_{n \to \infty} C_{\Gset,n} = 1$ and $\lim_{n \to \infty} r(n) = 0$.
(Recall that $C_n$ is given as \eqref{definition_C_n} and $C_n = 1+O(n^{-1/4})$ for $\beta = 1/4$.)
This result corresponds to \eqref{regret_exponential_family} for exponential families.

Under Assumption~\ref{assumption_boundary},
we can construct $\ptp'$ in \eqref{definition_combined_two_part_code} with $\ptp = \ptpbar$.
Let
\begin{align}
  \overline{\REG}
  =
  \alpha \pair[\Big]{
  \frac{K}{2} \log \frac{n}{2\pi} + \log \int_{\Theta} \abs{J(\theta)}^{1/2} d\theta
  + f_{\rm ne}(n) + l_1(n)
  }.
\end{align}
Then, we have the following corollary.
\begin{cor}[Bounds for non-exponential families]
  \label{cor_risk_bound_for_non_exponential_families}
  Let
  $\Mset$ is a non-exponential family which satisfies Assumptions~\ref{assumption_cont_differentiability}--\ref{assumption_V2}.
  When $p^* \in \Mset$,
  the MDL estimator induced by $\ptp'$ satisfies the following inequalities for large $n$.
  \begin{align}
    \label{ineq_BC_non_exp1}
    \E_{X^n \sim p^*} \dbar_\lambda(p^* \| \ddot{p}_{X^n}) \leq \frac{1}{n} \overline{\REG}.
  \end{align}
  and for all $b > 0$,
  \begin{align}
    \label{ineq_BC_non_exp2}
    \Pr \set[\Big]{\dbar_\lambda (p^* \| \ddot{p}_{X^n}) > \frac{1}{n} \overline{\REG} + b} < e^{-n \alpha^{-1} b},
  \end{align}
  where $\Pr$ is measured with $X^n \sim p^*$.
\end{cor}
We have obtained the bounds for non-exponential families which generalize the bounds for exponential families given by Corollary~\ref{BC_cor}.

In the rest of this section,
we will prove Theorem \ref{thm_improved_regret_bound}.
We need some definitions and lemmas to prove the theorem.

Let $\dabs{V}_s$ denote
the spectral norm of a real-symmetric matrix $V$:
\begin{align}
  \dabs{V}_s = \max_{\abs{z} = 1} \abs{z^T V z}.
\end{align}
The following lemma provides inequalities between the maximum norm and the spectral norm.
\begin{lemma}
  \label{lemma_ineq_for_norms}
  For an arbitrary $K \times K$ real-symmetric matrix $V$, the following inequality holds.
  \begin{align}
    \label{ineq_for_norms1}
    \frac{1}{\sqrt{K}} \dabs{V}_M \leq \dabs{V}_s \leq K \dabs{V}_M.
  \end{align}
\end{lemma}
\begin{IEEEproof}[Proof of Lemma~\ref{lemma_ineq_for_norms}]
  To prove \eqref{ineq_for_norms1},
  we use the Frobenius norm defined as
  \begin{align}
    \dabs{V}_F^2 = \sum_{i,j} V_{ij}^2.
  \end{align}

  By definition of $\dabs{V}_M$,
  $\dabs{V}_F^2 \leq K^2 \dabs{V}_M^2$ holds.
  Further, since
  $\dabs{V}_s^2$ is the maximum squared eigenvalue of $V$, and since
  $\dabs{V}_F^2$ $=\Tr (V^2)$ corresponds to the sum of the eigenvalues of $V^2$,
  which equals the sum of the squared eigenvalues of $V$,
  we have $\dabs{V}_s^2 \leq \dabs{V}_F^2$.
  By combining them, we have $\dabs{V}_s \leq K \dabs{V}_M$.
  In the same manner,
  $\dabs{V}_F^2 \leq K \dabs{V}_s^2$ and $\dabs{V}_M^2 \leq \dabs{V}_F^2$ holds.
  Therefore, we have $\dabs{V}_M \leq \sqrt{K} \dabs{V}_s$.
\end{IEEEproof}

We define a function $g(\theta, \xi; x^n)$, which appears in evaluation of the code length of our two part code, as follows:
\begin{align}
  g(\theta, \xi; x^n) = \xi \cdot V(\theta; x^n) - \psi_{\theta}(\xi).
\end{align}
The following lemma is used to provide an upper bound of the regret for $x^n \in \Gset^c$.
\begin{lemma}
  \label{lemma_g_bound}
  Let $\Delta$ be a constant $\Delta$ in Assumption~\ref{assumption_V1}.
  Suppose $x^n \in \Gset^c$ and $\abs{\thetaddh - \thetahat} < \Delta$.
  Then, under Assumptions~\ref{assumption_V1} and \ref{assumption_V2},
  for $\xi \in \Xi_n$ satisfying \eqref{selection_xi},
  \begin{align}
    g(\thetaddh, \xi; x^n) > u_n \dabs{V(\thetaddh; x^n)}_M \pair[\Big]{1 - \frac{B}{2\gamma \delta_n} u_n}.
  \end{align}
\end{lemma}
\begin{IEEEproof}[Proof of Lemma~\ref{lemma_g_bound}]
  When $\xi$ satisfies \eqref{selection_xi},
    \begin{align}
      g(\thetaddh, \xi; x^n) = u \dabs{V(\thetaddh; x^n)}_M - \psi_{\thetaddh}(\xi).
    \end{align}
  holds.
  First, we evaluate $\psi_{\thetaddh}(\xi)$.
  Suppose $\abs{\xi_{ij}} = u$ and $\xi_{i'j'} = 0$ for $(i', j') \neq (i, j)$.
  By the definition of $\psi_\theta$, we have
  \begin{align}
    \partv[\Big]{\frac{\partial \psi_{\theta}(\xi)}{\partial \xi_{ij}}}_{\xi = 0}
    =
    \E_{p_\theta} V_{ij}(\theta; X) = 0 
  \end{align}
  As for the second derivative, we have for any $v \in \Rset^{K\times K}$,
  \begin{align}
  \sum_{ijkl}v_{kl}v_{ij}
    \frac{\partial^2 \psi_{\theta}(\xi)}{\partial \xi_{kl}\partial \xi_{ij}}
    &=
     \sum_{ijkl}v_{kl}v_{ij}
    \E_{\pbar_{\theta, \xi}} V_{kl}(\theta; X)  V_{ij}(\theta; X) 
    -  \sum_{ijkl}v_{kl}v_{ij}E_{\pbar_{\theta, \xi}}V_{kl}(\theta; X)E_{\pbar_{\theta, \xi}}V_{ij}(\theta; X)\\
        &=
    \sum_{ijkl}v_{kl}v_{ij}
    \E_{\pbar_{\theta, \xi}} V_{kl}(\theta; X)  V_{ij}(\theta; X) 
    -  \pair[\Big]{\sum_{ij}v_{ij}E_{\pbar_{\theta, \xi}}V_{ij}(\theta; X)}^2\\
    &\leq  \sum_{ijkl}|v_{kl}v_{ij}| B.
  \end{align}
  Since $\psi_{\theta}(0) = \log \int p_\theta(x)dx= 0$,
  by Taylor expansion around $\xi_{ij} = 0$, we have for all $\theta$,
  \begin{align}
    \psi_{\theta}(\xi) \leq \frac{B}{2} \sum_{ijkl}\abs{\xi_{ij}\xi_{kl}} =  \frac{B}{2}u_n^2.
  \end{align}
  Therefore, we have
  \begin{align}
    g(\thetaddh, \xi; x^n)
    &\geq
    u \dabs{V(\thetaddh; x^n)}_M - \frac{B}{2} u_n^2\\
    &>
    u \dabs{V(\thetaddh; x^n)}_M \pair[\Big]{1 - \frac{B}{2 \gamma \delta_n} u_n}.
  \end{align}
\end{IEEEproof}

Now, we are ready to show Theorem \ref{thm_improved_regret_bound}.
\begin{IEEEproof}[Proof of Theorem \ref{thm_improved_regret_bound}]
Since we need the bound only for $x^n$ such that $\thetahat \in \interior{\Theta}$,
we suppose $\thetahat \in \interior{\Theta}$ in this proof.
By \eqref{bundlecode},
the code length of our code is
\begin{align}
  -\log \ptpbar(x^n)
  &= - \log p_{\ddot{\theta}}(x^n)
  - n \set[\Big]{\ddot{\xi} \cdot V(\ddot{\theta}; x^n) - \psi_{\ddot{\theta}}(\xi)} 
  + \alpha \Lbar_n(\thetaddot, \xiddot) \\
  &= - \log p_{\ddot{\theta}}(x^n) - n g(\ddot{\theta}, \ddot{\xi}; x^n) + \alpha \Lbar_n(\thetaddot, \xiddot) \\
   &\le - \log p_{\thetaddh}(x^n) - n g(\thetaddh, \bar{\xi}; x^n) + \alpha \Lbar_n(\thetaddh, \bar{\xi}),
\end{align}
where $\thetaddh$ is one of the closest point in $\ddotTheta$ to $\thetahat$
and $\bar{\xi}$ denotes the element of $\Xi_n$ which we will define 
depending on whether $x^n \in \Gset$ or not.
Hence, we have
\begin{align}\label{regret_expansion}
 \REG(\ptpbar; \Mset, x^n)=
  \log \frac{p_{\hat{\theta}}(x^n)}{\ptpbar(x^n)}
  \le \log \frac{p_{\hat{\theta}}(x^n)}{p_{\thetaddh}(x^n)} - n g(\thetaddh, \bar{\xi}; x^n) + \alpha \Lbar_n(\thetaddh, \bar{\xi}).
\end{align}
We are to give upper bounds on the right side of \eqref{regret_expansion}.

{\it Part I ($x^n \in \Gset$):}
First, we consider the case $x^n \in \Gset$,
for which we let $\bar{\xi}=0$.
Then, we have
\begin{align}\label{regret_expansion_for_good}
 \REG(\ptpbar; \Mset, x^n)
  \le \log \frac{p_{\hat{\theta}}(x^n)}{p_{\thetaddh}(x^n)}  
  + \alpha \Lbar_n(\thetaddh, 0 ).
\end{align}
By the same Taylor expansion as \eqref{regret_bound_by_Jhat_pre},
we have
\begin{align}\label{regret_expansion_for_good2}
 \REG(\ptpbar; \Mset, x^n)
  \le \frac{n}{2}(\thetaddh-\thetahat)^T\hat{J}(\theta';x^n) (\thetaddh-\thetahat) 
  + \alpha \Lbar_n(\thetaddh, 0 ),
\end{align}
where $\theta'$ is a point on the line segment between $\thetaddh$ and $\thetahat$.
Recall $\abs{\theta' - \thetahat}^2 \leq Ka^2/4n \zeta$.
(See the proof of Lemma~\ref{lemma_data_description_loss_bound}.)
Hence
by Assumption~\ref{assumption_epsilon_weak}, we have
\begin{align}\label{regret_expansion_for_good3}
 \REG(\ptpbar; \Mset, x^n)
  \le \frac{n}{2}(\thetaddh-\thetahat)^T\hat{J}(\thetahat;x^n) (\thetaddh-\thetahat) 
  (1+\kappa'K^{1/2}a \zeta^{-1/2}n^{-1/2}/2 )
  + \alpha \Lbar_n(\thetaddh, 0 ),
\end{align}
To evaluate the first term of the right side, 
we will prove 
\begin{align}
  z^T \Jhat(\theta; x^n) z \leq z^T J(\theta) z \pair[\Big]{1 + \dabs{V(\theta; x^n)}_s}
  \label{ineq_by_V2}
\end{align}
for all $z \in \Rset^K$ and all $\theta \in \Theta$, as follows.
By the definition of $V(\theta; x^n)$,
for all $z \in \Rset^K$,
\begin{align}
  (J^{1/2}(\theta) z)^T V(\theta; x^n) (J^{1/2}(\theta) z) = z^T \Jhat(\theta; x^n) z - z^T J(\theta) z
\end{align}
and
\begin{align}
  \abs[\Big]{J^{1/2}(\theta) z}^2 = z^T J(\theta) z
\end{align}
hold.
Hence
we have the following inequality for all $\theta$ and $x^n$.
\begin{align}
  \label{ineq_by_V}
  z^T \Jhat(\theta; x^n) z \leq z^T J(\theta) z \pair[\Big]{1 + \dabs{V(\theta; x^n)}_s},
\end{align}
which is \eqref{ineq_by_V2}.

Hence from \eqref{regret_expansion_for_good3}, we 
have
\begin{align}\label{regret_expansion_for_good4}
 \REG(\ptpbar; \Mset, x^n)
  \le \frac{n}{2}(\thetaddh-\thetahat)^T
  J(\thetahat) (\thetaddh-\thetahat) \pair[\Big]{1 + \dabs{V(\thetahat; x^n)}_s}
  (1+\kappa'K^{1/2}a \zeta^{-1/2}n^{-1/2}/2 )
  + \alpha \Lbar_n(\thetaddh, 0 ).
\end{align}
Then, by Lemma~\ref{lemma_data_description_loss_bound},
and since
$\dabs{V(\thetahat; x^n)}_s\le K\dabs{V(\thetahat; x^n)}_M \le K\delta_n$ 
holds for $x^n \in \Gset$,
we have
\begin{align}\label{regret_expansion_for_good5}
 \max_{x^n \in \Gset}
 \REG(\ptpbar; \Mset, x^n)
  &\le \frac{C_n Ka^2}{8}
  (1 + K\delta_n)
  (1+\kappa'K^{1/2}a \zeta^{-1/2}n^{-1/2}/2 )
  + \alpha \Lbar_n(\thetaddh, 0 )\\
  & \le \frac{C_{\Gset,n} Ka^2}{8}
  (1 + K\delta_n)
  + \alpha (L_n(\thetaddh) + l_2(n)), \label{thelastline}
\end{align}
where $C_{\Gset,n}$ is defined as \eqref{definition_C_gn}.
Recalling $\Lbar_n(\thetaddh, 0 )=L_n(\thetaddh)+l_2(n)$
and
Lemma~\ref{prop_L_n}, \eqref{thelastline}
is bounded upper by $\alpha$ times
the right side of 
\eqref{ineq_improved_regret_bound}.
That is, we have the demanded upper bound for $x^n \in \Gset$.

{\it Part II ($x^n \in \Gset^c$):} 
When $x^n \in \Gset^c$ holds,
$\dabs{V(\thetahat; x^n)}_M$ is not bounded in general.
However, Lemma \ref{lemma_g_bound} allows us to achieve a smaller regret than the minimax value.

Let $\bar{\xi}$ be an element of $\Xi_n$ such that \eqref{selection_xi} holds.
Note that since \eqref{quantize_ineq1}
implies $\abs{\thetaddh - \thetahat}^2 \leq Ka^2 / 4 n \zeta$,
the condition $\abs{\thetaddh - \thetahat} < \Delta$ in Lemma \ref{lemma_g_bound} is satisfied when
\begin{align}
  \label{condition_n_for_Delta}
  n > \frac{Ka^2}{4 \zeta \Delta^2}.
\end{align}
Since we let $u_n = \gamma \delta_n / B$, 
for the second term of \eqref{regret_expansion},
by Lemma~\ref{lemma_g_bound},
\begin{align}\label{g_lower_bound}
-n  g(\thetaddh, \bar{\xi}; x^n) < -n\frac{\gamma \delta_n}{2B} \dabs{V(\thetaddh; x^n)}_M 
\end{align}
holds for $n$ satisfying \eqref{condition_n_for_Delta}.

Next, we evaluate the first term of the right side of \eqref{regret_expansion}.
By Taylor expansion, we have
\[
\log \frac{p_{\hat{\theta}}(x^n)}{p_{\thetaddh}(x^n)}
= \frac{n (\thetaddh - \hat{\theta})^T\hat{J}(\theta',x^n) (\thetaddh - \hat{\theta})}{2},
\]
where $\theta'$ is a point on the line between $\thetaddh$ and $\hat{\theta}$.
Let $\epsilon$ be so small
that Assumption~\ref{assumption_epsilon} holds.
Then, by Assumption~\ref{assumption_J_eigen_values},
for all unit vector $z$,
and 
for all $\theta$ such that $\abs{\theta - \hat{\theta}} < \epsilon$, 
the following holds
\begin{align}
\frac{z^T\Jhat(\theta; x^n)z}{ z^TJ(\thetahat)z} \le
    \frac{C_{\epsilon} \pair[\Big]{z^T\Jhat(\thetahat; x^n)z + z^TJ(\thetahat)z}}{z^T J(\thetahat)z},
\end{align}
which yields
\begin{align}
z^T\Jhat(\theta; x^n)z \le
 z^TJ(\thetahat)z
    C_{\epsilon} (1+ \dabs{V(\thetahat; x^n)}_s).
\end{align}
Now, assume
\begin{align}\label{n_condition_epsilon}
    n > \frac{Ka^2}{4\zeta \epsilon^2}.
\end{align}
Then, by Proposition~\ref{prop_quantization},
$\abs{\thetaddh-\thetahat}< \epsilon$ holds.
Hence, by
Lemma~\ref{lemma_data_description_loss_bound}, the first term of \eqref{regret_expansion}
is bounded as
\[
\log \frac{p_{\hat{\theta}}(x^n)}{p_{\thetaddh}(x^n)}
\le
\frac{C_n C_\epsilon Ka^2(1+ \dabs{V(\thetahat; x^n)}_s)}{ 8 } 
\le
\frac{C_n C_\epsilon Ka^2(1+ K\dabs{V(\thetahat; x^n)}_M)}{ 8 } 
\]

Therefore, together with the bound for the second term, we have
\begin{align}
  \log \frac{p_{\thetahat}(x^n)}{\pbar_{\thetaddh, \bar{\xi}}(x^n)}
  <
   \frac{C_n C_\epsilon Ka^2}{8} - \dabs{V(\thetahat; x^n)}_M
  \pair[\Big]{\frac{\gamma \delta_n}{2B} \frac{\dabs{V(\thetaddh; x^n)}_M}{\dabs{V(\thetahat; x^n)}_M} n -  \frac{C_n C_\epsilon Ka^2}{8}}.
  \label{regret_bound_Gc_ddh_0}
\end{align}
Under Assumption~\ref{assumption_V1} and \eqref{condition_n_for_Delta},
we have
\begin{align}
  \log \frac{p_{\thetahat}(x^n)}{\pbar_{\thetaddh, \bar{\xi}}(x^n)}
  <
  \frac{C_n C_\epsilon K a^2}{8} - \dabs{V(\thetahat; x^n)}_M
  \pair[\Big]{\frac{\gamma \delta_n}{2B} n - \frac{C_n C_\epsilon Ka^2}{8}}.
  \label{regret_bound_Gc_ddh_1}
\end{align}
When $n$ satisfies
\begin{align}
    \label{condition_n_gamma_delta}
    n > \frac{C_n C_\epsilon Ka^2}{4\gamma \delta_n},
\end{align}
the second term is negative.
Then when $n$ satisfies \eqref{condition_n_gamma_delta},
for all $x^n \in \Gset^c$, since $\dabs{V(\thetahat; x^n)} > \delta_n$,
we have
\begin{align}
  \log \frac{p_{\thetahat}(x^n)}{\pbar_{\thetaddh, \bar{\xi}}(x^n)}
  <
   \frac{C_n C_\epsilon Ka^2}{8} -
  \frac{\gamma \delta_n^2}{2B} n + \frac{C_n C_\epsilon Ka^2}{8} \delta_n.
  \label{regret_bound_Gc_ddh}
\end{align}
Now, when $n$ satisfies 
\eqref{n_condition_epsilon} and  \eqref{condition_n_for_Delta},
which are denoted by
\begin{align}
    \label{n_condition_epsilon_Delta}
    n > \frac{Ka^2}{4 \zeta} \max \set[\Big]{\frac{1}{\epsilon^2}, \frac{1}{\Delta^2}},
\end{align}
and 
satisfies \eqref{condition_n_gamma_delta},
we have the following upper bound of the regret.
\begin{align}
\max_{x^n \in \Gset^c}
  \REG(\ptpbar; \Mset, x^n) 
  &<
   \frac{C_n C_\epsilon Ka^2(1+\delta_n)}{8} -
  \frac{\gamma \delta_n^2}{2B} n 
  + \alpha (L_n(\thetaddh) + \bar{l}_2(n)).
  \label{regret_bound_Gc}
\end{align}

Lastly, we will give the condition under which
the bound \eqref{regret_bound_Gc} is smaller than \eqref{thelastline}.
The right side of \eqref{thelastline}
minus that of \eqref{regret_bound_Gc}
is
\begin{align}
  &  \frac{C_{\Gset,n} Ka^2(1 + K\delta_n)}{8} 
  + \alpha (L_n(\thetaddh) + l_2(n))
  - \frac{C_n C_\epsilon Ka^2(1+\delta_n)}{8} 
  +
  \frac{\gamma \delta_n^2}{2B} n 
  - \alpha (L_n(\thetaddh) + \bar{l}_2(n))\\
  & >
  -\frac{C_n C_\epsilon Ka^2(1+\delta_n)}{8} 
  +\alpha (l_2(n) - \bar{l}_2(n))
   +
  \frac{\gamma \delta_n^2}{2B} n \\
  & >
  \frac{\gamma \delta_n^2}{2B} n 
  -\frac{C_n C_\epsilon Ka^2(1+\delta_n)}{8} 
  +\alpha (\log l_2(n) - \log (2K^2))\\
  & >
  \frac{\gamma \delta_n^2}{2B} n 
  -C_\epsilon Ka^2
  +\alpha (\log l_2(n) - \log (2K^2))\\
    & =
  \frac{\gamma g \log n}{2B} 
  -C_\epsilon Ka^2
  -\nu \alpha \log n
  -\alpha  \log (2K^2)\\
      & =
  \Bigl(\frac{\gamma g }{2B} -\nu \alpha \Bigr)\log n
  -C_\epsilon Ka^2
  -\alpha  \log (2K^2),
\end{align}
where the second inequality follows from \eqref{eq:defferenceL2}
and we assume $n$ satisfies $\delta_n \le 1$ and $C_n \le 4$.
Since $\gamma g/2B - \nu \alpha > 0$ is assumed, 
the above difference is positive, if
\[
\log n > 
\frac{C_\epsilon Ka^2  +\alpha  \log (2K^2)}{\gamma g/2B - \nu \alpha}
\]
holds.
Finally, we consider sufficient conditions for $\delta_n \le 1$ and $C_n \le 4$, respectively.
Since $x \ge \log 2x$ holds for $x \ge 1$,
we have $\log g \ge \log (2\log g) = \log 2 +\log \log g$ for $g \ge e$.
Further, $\log n /n$ decreases when $n \ge e$.
Therefore, if $n \ge 2 g \log g$, we have
\[
\delta_n^2 = \frac{g\log n}{n} 
\le 
\frac{\log \log g + \log (2 g)}{2 \log g}
\le \frac{2\log g}{2\log g}=1.
\]
As for $C_n$, since we assume $\beta = 1/4$ here,
when
\[
n \ge \max \Bigl\{ \frac{1}{\kappa^4 K^2 a^4}, \frac{4\zeta}{\kappa^2 K a^2}   \Bigr\}.
\]
Therefore, \eqref{condition_n0} is a sufficient condition
for \eqref{ineq_improved_regret_bound}.
We have obtained the claim of the theorem.
%
\end{IEEEproof}

\section{Proofs of a Proposition and Lemmas}
\label{section_proofs}

\begin{IEEEproof}[Proof for Proposition~\ref{prop_quantization}]
    Recall that the quantization is realized by the hyper rectangles which 
    circumscribe the ellipsoids defined as \eqref{definition_ellipsoid}.
    Note that the ellipsoid of scale $\sqrt{K}$ 
    circumscribes the rectangle which circumscribes the original rectangle.

    In a hyper rectangle induced by an ellipsoid \eqref{definition_ellipsoid},
    the points which are farthest from the center of mass $\theta$ of the hyper rectangle are on its corners.
    Let $\theta'$ be one of these corners.
    Since $\abs{\theta' - \theta}$ is not greater than $\sqrt{K}$ times of the largest length of the edges of the rectangle,
    all the corners are inside of the following ellipsoid with $\sqrt{K}$ times larger scale as 
    \begin{align}
      \set[\Big]{\theta' : (\theta' - \theta)^T J(\theta_{\Scell}) (\theta' - \theta) \leq \frac{K a^2}{4n}}.
    \end{align}
    Therefore, \eqref{quantize_ineq1} holds.
    Finally, \eqref{quantize_ineq2} can easily be derived as well. 
\end{IEEEproof}

\begin{IEEEproof}[Proof for Lemma~\ref{prop_L_n}]
  For each large cell $\Scell$, 
  let $\abs{\ddot{\Theta}_{n, \Scell}}$ denote the number of
  the small cells which intersect $\Scell$.
  We will evaluate its upper bound. 
  For a measurable bounded subset $E \subset \Rset^K$, we denote the volume of $E$ as
  \begin{align}
    \sigma(E) = \int_{E} d\theta.
  \end{align}

  Let $\bar{\Scell}$ denote the union of all the hyper rectangles which intersect $\Scell$, and $\sigma_{\rm hr}$ 
  the volume of the hyper rectangles.
  Note that $\Scell \subseteq \bar{\Scell}$
  and that $\sigma_{\rm hr}=(a n^{-1/2})^K \abs{J(\theta_{\Scell})}^{-1/2}$.
  Since each hyper rectangle includes just one quantized point,
  we have
  \begin{align}
    \abs{\ddot{\Theta}_{n, \Scell}} = \frac{\sigma(\bar{\Scell})}
    {\sigma_{\rm hr}}.
    \label{eq_ddot_Theta_n_Scell}
  \end{align}

  We will give an upper bound of $\sigma(\bar{\Scell})$.
  Let $N_{\Scell}$ be the number of the small cells which intersect the surface of $\Scell$.
  First, we evaluate an upper bound of $N_{\Scell}$.
  Recall that $\Scell$ is the intersection of the convex set $\Theta$
  and a hyper cube (with side lengths $a n^{-\beta}$), which we will denote by $C(\Scell)$ here. 
  Note that
  the number of the hyper rectangles to cover the surface of $C(\Scell)$ is larger than $N_{\Scell}$.
  When the number is maximized, the smallest edge of the rectangle is parallel to an edge of the hyper cube $C(\Scell)$.
  Further the number is upper bounded by that for the case in which the lengths of all the edges are equal to the minimum value.
  Since, from Assumption~\ref{assumption_J_eigen_values},
  the maximum of the eigenvalue of $J(\theta_{\Scell})$ is at most $\lambdabar$,
  the length of the shortest edge of the rectangle is at least 
  $a n^{-1/2} \lambdabar^{-1/2}$.
  Then, we have
  \begin{align}
    N_{\Scell}
    &\leq 2K
    \Bigl(
    \frac{a n^{-\beta} }{ an^{-1/2} \lambdabar^{-1/2}} + 2
    \Bigr)^{K - 1} \\
    &\leq 2K n^{(K - 1)(1/2-\beta)} (\lambdabar^{1/2} + 2)^{K - 1},
  \end{align}
  where the factor $2K$ is 
  the number of the $K-1$ dimensional hyper cubes which
  form the surface of $C(\Scell)$.
  Let $\bar{N}$,
  which is independent of $\Scell$, denote the last side of the above inequalities.
  Then, we have
  \begin{align}
    \abs{\ddot{\Theta}_{n, \Scell}}
    \le \frac{\sigma(\Scell)}{\sigma_{\rm hr}} + \bar{N}.
  \end{align}
  Since $\Scell$ is a subset of a hyper cube with side lengths $an^{-\beta}$,
 $\sigma(\Scell) \leq (an^{-\beta})^K$ holds.
 Further, recall that $\sigma_{\rm hr}=(a n^{-1/2})^K \abs{J(\theta_{\Scell})}^{-1/2}$. Then, we have
 \begin{align}
    \abs{\ddot{\Theta}_{n, \Scell}}
    \le 
    \frac{(an^{-\beta})^K}{(a n^{-1/2})^K \abs{J(\theta_{\Scell})}^{-1/2}} 
    + \bar{N}
    =
    n^{K(1/2-\beta)}  \abs{J(\theta_{\Scell})}^{1/2}+ \bar{N}.
  \end{align}

  Therefore, we have
  \begin{align}
    \sigma(\bar{\Scell})
    &\leq
    \sigma(\Scell) + \bar{N} (a n^{-1/2})^K \abs{J(\theta_{\Scell})}^{1/2} \\
    &\leq
    \sigma(\Scell) + 2K n^{(K - 1)(1/2-\beta)} (\lambdabar^{1/2} + 2)^{K - 1}(a n^{-1/2})^K \abs{J(\theta_{\Scell})}^{1/2} \\
    &\leq
    \sigma(\Scell) + a^K C_J n^{- K\beta -(1/2-\beta)},
    \label{ineq_sigma_bar_Scell1}
  \end{align}
  where
  \begin{align}
    \label{definition_C_J}
    C_J = 2K(\lambdabar^{1/2} + 2)^{K - 1} \Lambda^{1/2}.
  \end{align}
  Under Assumptions~\ref{assumption_J_eigen_values} and \ref{assumption_J_determinant},
  \begin{align}
    \label{definition_Lambda}
    \Lambda = \max_{\theta \in \Theta} \abs{J(\theta)}.
  \end{align}
  is finite.

  Since $\Scell$ is a subset of the hyper cube with side lengths $an^{-\beta}$, which induces it,
  $\sigma(\Scell) \leq (an^{-\beta})^K$ holds.
  Then,
  \begin{align}
    \sigma(\bar{\Scell})
    \leq
    (an^{-\beta})^K (1 + C_J n^{-(1/2-\beta)})
    \label{ineq_sigma_bar_Scell_general}
  \end{align}
  holds.

  When $\Scell \cap \boundary{\Theta} = \emptyset$,
  since
  $\Scell$ is a hyper cube,
  $\sigma(\Scell) = (a n^{-\beta})^K$ holds.
  We have
  \begin{align}
    \sigma(\bar{\Scell})
    \leq
    \sigma(\Scell)(1 + C_J n^{-(1/2-\beta)}).
    \label{ineq_sigma_bar_Scell2}
  \end{align}
  Now, we can derive the following inequality under Assumption~\ref{assumption_D_J}.
  \begin{align}
    \sigma(\Scell)
    \leq
    \frac{\int_{\Scell} \abs{J(\theta)}^{1/2} d\theta}{\abs{J(\theta_{\Scell})}^{1/2}} (1 + C_{\Theta} n^{-\beta}),
    \label{ineq_sigma_Scell}
  \end{align}
  where
  \begin{align}
    \label{definition_C_Theta}
    C_{\Theta} = \max_{\theta \in \Theta} \frac{K a D_J}{\abs{J(\theta)}^{1/2}}.
  \end{align}
  In fact, for all $\theta \in \Scell$,
  by Taylor expansion, we have
  \begin{align}
    \abs{J(\theta_{\Scell})}^{1/2} \leq \abs{J(\theta)}^{1/2} + K a D_J n^{-\beta}.
  \end{align}
  That is,
  \begin{align}
    1 \leq \frac{\abs{J(\theta)}^{1/2}}{\abs{J(\theta_{\Scell})}^{1/2}} (1 + C_{\Theta} n^{-\beta}).
  \end{align}
  By integrating both sides, we have \eqref{ineq_sigma_Scell}.

  Finally, by using \eqref{eq_ddot_Theta_n_Scell}, \eqref{ineq_sigma_bar_Scell2} and \eqref{ineq_sigma_Scell},
  we have for $\Scell$ such that $\Scell \cap \boundary{\Theta} = \emptyset$,
  \begin{align}
    \abs{\ddotTheta_{n, \Scell}}
    \leq
    a^{-K} n^{K/2} (1 + C_J n^{-(1/2-\beta)}) (1 + C_{\Theta} n^{-\beta}) \int_{\Scell} \abs{J(\theta)}^{1/2} d\theta.
  \end{align}
  Let $\Theta_{\partial}$ denote the union of large cells which have intersections with $\boundary{\Theta}$.
  By taking the summation over all large cells without intersections with $\boundary{\Theta}$,
  an upper bound of the number of small cells induced by such large cells is evaluated as
  \begin{align}
    &a^{-K} n^{K/2} (1 + C_J n^{-(1/2-\beta)}) (1 + C_{\Theta} n^{-\beta}) \int_{\Theta \setminus \Theta_{\partial}} \abs{J(\theta)}^{1/2} d\theta \\
    &\leq
    a^{-K} n^{K/2} (1 + C_J n^{-(1/2-\beta)}) (1 + C_{\Theta} n^{-\beta}) \int_{\Theta} \abs{J(\theta)}^{1/2} d\theta.
    \label{number_of_quantized_points_avoiding_boundary}
  \end{align}

  Next, we consider $\Scell$ (large cell) with $\Scell \cap \boundary{\Theta} \neq \emptyset$.
  Define a real number $W_\Theta$ as
  \[
W_\Theta = \max_i \bigl( \max\{\theta_i : \theta \in \Theta  \} -\min\{ \theta_i : \theta \in \Theta \}
  \bigr).
  \]
  Since $\Theta$ is included in the hyper cube with side $W_\Theta$,
  the number of large cells (hyper cubes) which have intersections with $\boundary{\Theta}$ is at most
  \begin{align}
    2^K \pair{W_{\Theta} a^{-1} n^{\beta} + 2}^{K - 1}
    \leq
    C_K a^{-(K - 1)} n^{(K - 1)\beta},
    \label{number_of_cubes_boundary}
  \end{align}
  where
  \begin{align}
    \label{definition_C_K}
    C_K = 2^K (W_{\Theta} + 2a)^{K - 1}.
  \end{align}
  By using \eqref{eq_ddot_Theta_n_Scell} and \eqref{ineq_sigma_bar_Scell_general},
  we have for all large cells,
  \begin{align}
    \abs{\ddot{\Theta}_{n, \Scell}} \leq (1 + C_J n^{-(1/2-\beta)}) \Lambda^{1/2} n^{K (1/2-\beta)}.
  \end{align}
  By taking the summation of $\abs{\ddot{\Theta}_{n, \Scell}}$ over the hyper cubes which constitutes $\Theta_{\partial}$,
  since the number of them is at most the right side of \eqref{number_of_cubes_boundary},
  the number of small cells induced by them is at most
  \begin{align}
    C_K \Lambda^{1/2} a^{-(K - 1)} n^{K/2 - \beta} (1 + C_J n^{-(1/2-\beta)}).
  \end{align}
  By adding this to \eqref{number_of_quantized_points_avoiding_boundary},
  since $1 + C_J n^{-(1/2-\beta)}$ and $1 + C_{\Theta} n^{-\beta}$ are greater than $1$,
  we have
  \begin{align}
    \abs{\ddotTheta_{n}}
    \leq
    a^{-K} n^{K/2} e^{r(n)} \int_{\Theta} \abs{J(\theta)}^{1/2} d\theta,
  \end{align}
  where
  \begin{align}
    r(n) =
    \log (1 \!+\! C_J n^{-(1/2-\beta)}) + \log (1 \!+\! C_{\Theta} n^{-\beta}) + \log (1+C_{J, K} n^{-\beta})
    \label{definition_r}
  \end{align}
  and
  \begin{align}
    \label{definition_C_JK}
    C_{J, K} = \frac{C_K \Lambda^{1/2} a}{\int_{\Theta} \abs{J(\theta)}^{1/2} d\theta}.
  \end{align}
\end{IEEEproof}

\section{Application to Mixture Families}
In this section,
we will apply our result to mixture families which are an example of non-exponential families.

For $i = 0, 1, 2, \cdots, K$,
let $q_i$ be a probability density function on $\Xset$.
The mixture family which consists of $\set{q_i}$ is the family of convex combinations of $\set{q_i}$ as follows.
For $\tau \in [0, 1/(K + 1)]$, we define
the mixture family $\Mset_{\tau}$ by
\begin{align}
  p_\theta(x) &= \sum_{i=0}^{K} \theta_i q_i(x),\\
  \Theta & = \Theta_{\tau} = \set{\theta \in [0, 1]^{K} : \theta_i \geq \tau ~\text{for each $i \in \set{0, 1, \cdots, K}$}},\\
   \Mset & = \Mset_{\tau} = \set{p_\theta : \theta \in \Theta_{\tau}},
\end{align}
where $\theta = (\theta_1,\ldots,\theta_K)$ and
$\theta_0 = 1 - \sum_{i=1}^{K} \theta_i$.
In this paper, we fix $\tau > 0$.
Note that the naturally determined parameter space $\tilde{\Theta}$
equals $\Theta_\tau$ with $\tau = 0$.
In this paper, we consider the case that $\tau > 0$.
Note that the set $\{\theta \in \Theta : \theta_0=\tau \}$
is a subset of $\partial \Theta$.

Further, we assume that
each $q_i$ cannot be represented as a convex 
combination of the others, that is
for all $i \geq 0$,
\begin{align}
  \min_{(\alpha_0, \alpha_1, \ldots, \alpha_K) \in [0,1]^{K+1}
  :
  \sum_{j \neq i} \alpha_j = 1}
  D\Bigl(q_i \Big|\Big| \sum_{j \neq i} \alpha_j q_j \Bigr) > 0.
  \label{assumption_mixture_components_independence}
\end{align}
This condition guarantees that the Fisher information 
is positive definite.

We will show some properties of the mixture family.
The empirical Fisher information of the mixture family is given as follows.
\begin{align}
  \label{Jhat_of_mixture_family}
  \Jhat_{ij}(\theta; x) = \frac{(q_i(x) - q_0(x))(q_j(x) - q_0(x))}{p_{\theta}^2(x)}.
\end{align}
The following Lemma holds for mixture families in general.
\begin{lemma}
  \label{lemma_mixture_semi_positive}
  The empirical Fisher information of the mixture family is positive semi-definite.
  That is, for all unit vectors $z \in \Rset^K \setminus \set{0}$ and for all $x$,
  \begin{align}
    z^T \Jhat(\theta; x) z \geq 0.
  \end{align}
\end{lemma}
\begin{IEEEproof}
  From \eqref{Jhat_of_mixture_family}, we have
  \begin{align}
    z^T \Jhat(\theta; x) z
    &=
    \sum_{i,j} \pair[\Big]{\frac{q_i(x) - q_0(x)}{p_{\theta}(x)}} \pair[\Big]{\frac{q_j(x) - q_0(x)}{p_{\theta}(x)}} z_i z_j\\
    &=
    \pair[\Big]{\sum_{i} \frac{q_i(x) - q_0(x)}{p_{\theta}(x)} z_i}^2
    \geq 0.
    \label{lower_bound_of_Jhat_for_mixture_families}
  \end{align}
\end{IEEEproof}
We can show the positive-definiteness of $J(\theta)$.
Therefore, the assumption of existence of $\zeta$ in Assumption~\ref{assumption_J_eigen_values} is valid for mixture families.
\begin{lemma}
  \label{lemma_mixture_positive_definite}
  When $q_0, q_1, \cdots, q_K$ satisfy the condition \eqref{assumption_mixture_components_independence},
  for all $\theta$, $J(\theta)$ is positive-definite.
\end{lemma}
\begin{IEEEproof}
  From Lemma~\ref{lemma_mixture_semi_positive},
  $z^T J(\theta) z \geq 0$ holds for all $z \neq 0$.
  Then,
  it is sufficient to show
  $z^T J(\theta) z \neq 0$ for all $z \neq 0$.

  If $z^T J(\theta) z = 0$ for a certain $z \neq 0$,
  then $z^T \Jhat(\theta; x) z = 0$ must hold almost surely with respect to $p_{\theta}$.
  It imples
  \[
  \sum_{i} \frac{q_i(x) - q_0(x)}{p_{\theta}(x)} z_i = 0
  \]
  because of \eqref{lower_bound_of_Jhat_for_mixture_families}.
  Hence
  \begin{align}
      \label{char_equation_2}
      \sum_{j=1}^K (q_j(x) - q_0(x)) z_j = 0
  \end{align}
  must hold almost surely with respect to each of $q_0, q_1, \cdots, q_K$.
  (Recall $\theta_i > 0$ for all $i \ge 0$.)
  Since $z \neq 0$,
  there exists $i > 0$ such that $z_i \neq 0$.
  Let $z_0=-\sum_{i=1}^K z_i$. Then
  \[
   \sum_{i=0}^K z_i q_i(x)= 0
  \]
  holds almost surely with respect to each of $q_0, q_1, \cdots, q_K$.
  Note that $z_j\neq 0$ for at least one $j$. Let $b$ denote such $j$.
  Then, we have
  \[
  q_{b}(x)=-\sum_{i \neq b} (z_i/z_b) q_i(x).
  \]
  Since this holds almost surely with respect to $q_b$,
  \begin{align}
      D(q_b \| \sum_{i \neq b} (z_i/z_b) q_i) = 0
  \end{align}
  holds.
  This contradicts the condition \eqref{assumption_mixture_components_independence}.
  Therefore, $z^T J(\theta) z \neq 0$.
\end{IEEEproof}

For $\theta \in \Theta_{\tau}$, since $\theta_i \geq \tau$ for all $i$,
we have
\begin{align}
  \abs[\Big]{\frac{q_i(x) - q_0(x)}{p_{\theta}(x)}} \leq \frac{1}{\tau}.
\end{align}
Therefore,
\begin{align}
  \label{Jhat_element_bound}
  \Jhat_{ij}(\theta; x) \leq \frac{1}{\tau^2}
\end{align}
holds for all $\theta \in \Theta_{\tau}$, $x, i$ and $j$.
This implies that $\dabs{J(\theta)}_M \leq 1 / \tau^2$ for all $\theta \in \Theta_{\tau}$.
Then, from \eqref{ineq_for_norms1},
we have for all unit vectors $z \in \Rset^K$,
\begin{align}
  \max_{\theta \in \Theta_{\tau}} z^T J(\theta) z \leq \frac{K}{\tau^2}.
\end{align}
Therefore, Assumption~\ref{assumption_J_eigen_values} is valid for 
the mixture family $\Mset_{\tau}$
with
\begin{align}
  \lambdabar = \frac{K}{\tau^2}.
\end{align}
This gives an upper bound on $\Lambda$ defined by \eqref{definition_Lambda},
\begin{align}
  \Lambda \leq \lambdabar^K = \pair[\Big]{\frac{K}{\tau^2}}^{K}.
\end{align}

Further, since
\begin{align}
  \frac{\partial}{\partial \theta_h} \Jhat_{ij}(\theta; x) = - 2 \frac{q_h(x) - q_0(x)}{p_{\theta}(x)} \Jhat_{ij}(\theta; x),
  \label{mixture_bound_for_derivative}
\end{align}
we have for all $z \in \Rset^K$,
\begin{align}
  \label{mixture_Jhat_deriv_ineq}
  \abs[\Big]{\frac{\partial}{\partial \theta_h} z^T \Jhat(\theta; x) z}
  \leq \frac{2}{\tau} z^T \Jhat(\theta; x) z.
\end{align}
From Lebesgue's dominated convergence theorem, \eqref{mixture_Jhat_deriv_ineq}
implies
\begin{align}
    \label{mixture_J_deriv_ineq}
    \abs[\Big]{\frac{\partial}{\partial \theta_h} z^T J(\theta) z}
    \leq
    \frac{2}{\tau} z^T J(\theta) z.
\end{align}
From this, we can show the following lemma
which implies that
Assumption~\ref{assumption_kappa} holds with
\begin{align}
    B = \frac{\tau}{2 \sqrt{K}}
\end{align}
and
\begin{align}
    \label{value_kappa_for_mixture}
    \kappa = \frac{2e}{\tau} \sqrt{K}.
\end{align}.
\begin{lemma}
  \label{lemma_mixture_assumption_kappa}
  For for all $z \neq 0$,
  when $\theta_1, \theta_2 \in \Theta_\tau$
  satisfies
  $\abs{\theta_1 - \theta_2} \leq \tau / 2\sqrt{K}$,
  \begin{align}
    \label{ineq_mixture_assumption_kappa}
      \frac{z^T J(\theta_1) z}{z^T J(\theta_2) z}
      \leq
      1 + \frac{2e}{\tau} \sqrt{K} \abs{\theta_1 - \theta_2}
  \end{align}
  holds.
\end{lemma}
\begin{IEEEproof}
Since $z^T J(\theta) z > 0$,
from \eqref{mixture_Jhat_deriv_ineq}, we have
\begin{align}
    \frac{\partial}{\partial \theta_h}
    \log z^T J(\theta)z
    \leq
    \frac{2}{\tau}
\end{align}
for all $h$, $\theta \in \Theta_\tau$ and $z \neq 0$.
From Taylor expansion,
for all $\theta_1, \theta_2 \in \Theta_\tau$ and $z \neq 0$,
there exists $\theta'$ between $\theta_1$ and $\theta_2$ such that
\begin{align}
    \log z^T J(\theta_1)z
    =
    \log z^T J(\theta_2) z + 
    \sum_h \partv[\Big]{
        \frac{\partial}{\partial \theta_h} \log z^T J(\theta) z
        }_{\theta = \theta'}
    (\theta_1 - \theta_2)_h,
\end{align}
where $(\theta_1 - \theta_2)_h$ denotes the $h$-th element of $\theta_1 - \theta_2$.
Then from \eqref{mixture_Jhat_deriv_ineq}, we have
\begin{align}
    \log z^T J(\theta_1) z
    &\leq
    \log z^T J(\theta_2) z
    +
    \frac{2}{\tau} \sum_h \abs{(\theta_1 - \theta_2)_h} \\
    &\leq
    \log z^T J(\theta_2) z
    +
    \frac{2}{\tau} \sqrt{K} \abs{\theta_1 - \theta_2}.
\end{align}
Therefore,
\begin{align}
    \frac{z^T J(\theta_1) z}{z^T J(\theta_2) z}
    &\leq
    e^{2 \sqrt{K} \abs{\theta_1 - \theta_2}/\tau} \\
    &\leq
    1 + \frac{2\sqrt{K}\abs{\theta_1 - \theta_2}}{\tau}  e^{2 \sqrt{K} \abs{\theta_1 - \theta_2}/\tau} 
\end{align}
holds. Here, we have used the inequality $e^t \le 1+te^t$ derived from $e^{-t}\ge 1-t$.
When
$\abs{\theta_1 - \theta_2} \leq \tau / 2\sqrt{K}$,
the second term is not larger than
$2 e\sqrt{K} \abs{\theta_1 - \theta_2}/\tau$.
Hence we have the inequality \eqref{ineq_mixture_assumption_kappa}.
\end{IEEEproof}
The equation \eqref{mixture_bound_for_derivative}
leads to the following Lemma, which implies that
Assumptions \ref{assumption_epsilon_weak}\ and \ref{assumption_epsilon} are satisfied
with $B' = \tau / 2\sqrt{K} (=B)$ and $\kappa' = 2e\sqrt{K}/\tau = \kappa$.
\begin{lemma}
\label{lemma_mixture_assumption_epsilon2}
  For all $z$,
  when $\theta \in \Theta_\tau$
  satisfies
  $\abs{\theta - \hat{\theta}} \leq \tau / 2\sqrt{K}$,
  \begin{align}
    \label{ineq_mixture_assumption_kappa2}
      {z^T \hat{J}(\theta_1,x^n) z}
      \leq
      \Bigl(1 + \frac{2e\sqrt{K} \abs{\theta - \hat{\theta}} }{\tau }\Bigr)
      {z^T \hat{J}(\hat{\theta},x^n) z}
  \end{align}
  holds.
  Further, for an arbitrary constant $\epsilon > 0$, for all $x^n$ and for all $\theta$ such that $\abs{\theta - \thetahat} < \epsilon$,
  \begin{align}
    C_{\epsilon} = e^{{2} \sqrt{K} \epsilon/{\tau}}
    \label{value_C_epsilon_for_mixutre_families}
  \end{align}
  satisfies the following inequality for all unit vector $z \in \Rset^K$,
  \begin{align}
  \label{ineq_mixture_assumption_epsilon}
    z^T \Jhat(\theta; x^n) z
    \leq
    C_{\epsilon} z^T \Jhat(\thetahat; x^n) z .
  \end{align}
\end{lemma}
\begin{IEEEproof}
  From \eqref{mixture_bound_for_derivative},
  \begin{align}
    \abs[\Big]{
    \frac{\partial}{\partial \theta_h} \pair[\Big]{z^T \Jhat(\theta; x^n) z + s}
    } 
    \leq
    \frac{2}{\tau} \pair[\Big]{z^T \Jhat(\theta; x^n) z + s}
  \end{align}
  holds for any positive number $s$.
  Since $z^T \Jhat(\theta; x^n) z + s > 0$ for any $\theta \in \Theta_\tau$,
  by the same way as Lemma~\ref{lemma_mixture_assumption_kappa}, we have
  \begin{align}
    z^T \Jhat(\theta; x^n) z + s
    &\leq
    e^{2\sqrt{K} \abs{\theta - \thetahat}/\tau} \pair[\Big]{z^T \Jhat(\thetahat; x^n) z + s},
  \end{align}
  which is
   \begin{align}
    z^T \Jhat(\theta; x^n) z 
    &\leq
    e^{2\sqrt{K} \abs{\theta - \thetahat}/\tau} 
    z^T \Jhat(\thetahat; x^n) z 
    + (e^{2\sqrt{K} \abs{\theta - \thetahat}/\tau} -1)s,
  \end{align} 
  Since $e^{2\sqrt{K} \abs{\theta - \thetahat}/\tau}$ is bounded
  and $s$ can be arbitrarily small, we have
    \begin{align}\label{emp_ineq}
    z^T \Jhat(\theta; x^n) z 
    &\leq
    e^{2\sqrt{K} \abs{\theta - \thetahat}/\tau}
    z^T \Jhat(\thetahat; x^n) z .
  \end{align} 
  Hence, similarly as the proof of Lemma~\ref{value_kappa_for_mixture}, we 
  can obtain \eqref{ineq_mixture_assumption_kappa2}.
  Derivation of the inequality \eqref{ineq_mixture_assumption_epsilon} is straightforward from \eqref{emp_ineq}.
\end{IEEEproof}

From Assumption~\ref{assumption_J_eigen_values}, for all $\theta$,
\begin{align}
    \dabs{J^{-1/2}(\theta)}_s \leq \frac{1}{\sqrt{\zeta}}
\end{align}
holds
and hence, we have
\begin{align}
    \dabs{J^{-1/2}(\theta)}_M \leq \sqrt{ \frac{K}{\zeta}}.
\end{align}
In addition to this, derivatives of $\Jhat(\theta; x^n)$ and $J(\theta)$ is bounded.
Then, the derivative of $V(\theta; x^n)$ is also bounded.
Therefore, Assumption~\ref{assumption_V1} holds.
The following lemma implies that Assumption~\ref{assumption_V2} is satisfied.
\begin{lemma}
  For mixture family $\Mset_{\tau}$, for all $\theta \in \Theta_\tau$ and $x$,
  the following holds.
  \begin{align}
    \label{V_bound_for_mixture_family}
    \dabs{V(\theta; x)}_M \leq \frac{K \sqrt{K}}{\zeta \tau^2} + 1.
  \end{align}
\end{lemma}
\begin{IEEEproof}
  For all unit vectors $z \in \Rset^K$,
  \begin{align}
    z^T (J^{-1/2}(\theta) \Jhat(\theta; x) J^{-1/2}(\theta)) z
    &\leq
    \dabs{\Jhat(\theta; x)}_s \abs{J^{-1/2}(\theta) z}^2 \\
    &=
    \dabs{\Jhat(\theta; x)}_s z^T J^{-1}(\theta) z \\
    &\leq
    \dabs{\Jhat(\theta; x)}_s \dabs{J^{-1}(\theta)}_s
  \end{align}
  holds.
  From the second inequality of \eqref{ineq_for_norms1} and \eqref{Jhat_element_bound},
  $\dabs{\Jhat(\theta; x)}_s \leq K \dabs{\Jhat(\theta; x)}_M \leq K/\tau^2$ holds.
  Since eigenvalues of $J^{-1}(\theta)$ are inverses of eigenvalues of $J(\theta)$,
  we have $\dabs{J^{-1}(\theta)}_s \leq 1/\zeta$.
  Therefore, by using the first inequality of \eqref{ineq_for_norms1},
  we have
  \begin{align}
    \dabs{J^{-1/2}(\theta) \Jhat(\theta; x) J^{-1/2}(\theta)}_M
    \leq
    \frac{K \sqrt{K}}{\zeta \tau^2}.
  \end{align}
  By adding the unit matrix $I$, we have \eqref{V_bound_for_mixture_family}.
\end{IEEEproof}

Therefore, we can apply Theorem~\ref{thm_improved_regret_bound} to the mixture family.
\begin{thm}
  For the mixture family, we can construct a two part code for
  $x^n$ with $\hat{\theta} \in \Theta^\circ$
  whose regret is bounded by \eqref{ineq_improved_regret_bound}.
\end{thm}

To obtain the risk bounds as in Corollary~2, 
we have to care the sequences $x^n$ with $\hat{\theta} \in \partial \Theta$,
that is, we have to show the target model satisfies Assumption \ref{assumption_boundary}.
In fact,
we can construct an extra two part code in Assumption \ref{assumption_boundary}.
When $\thetahat \in \boundary{\Theta}$,
some elements of $\thetahat$ equal $\tau$.
First, consider a special case that
$\hat{\theta}_i$ for $i= 0, 1, 2, \ldots, K'$ are larger than $\tau$
and the others equal $\tau$.
Define
\begin{align*}
    \tilde{\Theta}' & = \{ \theta \in \Theta_0 : \theta_{K'+1}=\theta_{K'+2}= \cdots =\theta_K=\tau \},\\
       \Theta' & = \{ \theta \in \Theta_\tau : \theta_{K'+1}=\theta_{K'+2}= \cdots =\theta_K=\tau \},\\ 
    {\Theta'}^\circ & = \{ \theta \in \Theta_\tau : 
    \forall i \le K', \theta_i > \tau \text{ and }
    \theta_{K'+1}=\theta_{K'+2}= \cdots =\theta_K=\tau \}.
\end{align*}
Then, ${\Theta'}^\circ \subset \Theta' \subset \tilde{\Theta}'$ holds and
the special case here corresponds to the case that  $\hat{\theta}(x^n) \in \Theta'^\circ $.
Note that the set of densities
$
\{ p_\theta : \theta \in \tilde{\Theta}' \}
$
is the set of all the convex combinations of
\[
q'_i = \bigl(1-\tau (K-K') \bigr)q_i + \tau \sum_{j=K'+1}^K q_i
\]
for $i \in \{ 0,1,\ldots,K' \}$, which
is the mixture family based on the set of probability densities
$\{ q'_i : 0 \le i \le K'\}$.
Further,
$
\{ p_\theta : \theta \in \Theta' \}
$
can be a target model for Theorem~3 with the dimension $K'$.
Therefore, we can construct a two-part code whose regret
for the set $\{ x^n : \hat{\theta} \in \Theta'^\circ \}$
is not larger than 
\[
\frac{K'}{2}\log n + o(\log n).
\]
In this special case, though we fix the elements of $\theta$ which has the value larger than $\tau$,
there are $C(K,K')$ cases for each $K' \in \{ k : 0 \le k \le K -1  \}$, 
where $C(K,K')$ stands for a binomial coefficient.
The case that $\hat{\theta} \in \partial \Theta$
is included in one of the above cases.
Description of the value $K'$ and the choice of $\theta$'s elements larger than $\tau$
require extra code length to describe which case applies, but
such code length does not depend on $n$.
Therefore, the worst case regret for the case that $\hat{\theta} \in \partial \Theta$
is 
\[
\frac{K-1}{2}\log n + o(\log n).
\]
Hence, Assumption~6 with $d = 1$ holds for the mixture family $\Mset_\tau$
and we have the following theorem.
\begin{thm}
  \label{thm_for_mixture_families}
  For mixture families,
  the two part code constructed in this section induces an MDL estimator
  which satisfies the followings.
  For all $p^* \in \Mset$ and large $n$,
  \begin{align}
    \E_{X^n \sim p^*} \dbar_\lambda(p^* \| \ddot{p}_{X^n}) \leq \frac{1}{n} \overline{\REG}
  \end{align}
  and
  \begin{align}
    \Pr \set[\Big]{\dbar_\lambda (p^* \| \ddot{p}_{X^n}) > \frac{1}{n} \overline{\REG} + b} < e^{-n \alpha^{-1} b}
  \end{align}
  hold, where
  \begin{align}
    \frac{1}{\alpha} \overline{\REG} 
    =
      \frac{K}{2} \log n + \log \int_{\Theta} \abs{J(\theta')}^{1/2} d\theta' 
    +
      \frac{C_{\Gset,n}Ka^2}{8 \alpha} (1 + K\delta_n) + r(n) - K\log a+ c_0 + c_n
  \end{align}
  and
  $c_0, c' > 0$ is small constants used in the coding, which we can select arbitrarily.
  For the definitions of $C_{\Gset,n}$ and $r(n)$, see
  \eqref{definition_C_n}, 
  \eqref{definition_C_gn}
  and \eqref{definition_r}.
\end{thm}
Note that
$\lim_{n \to \infty} r(n) = 0$ and $\lim_{n \to \infty} C_{\Gset,n} = 1$
and
$C_{\epsilon} = e^{\frac{2}{\tau} \sqrt{K} \epsilon}$.

We will finish this section with evaluation of values or upper bounds of some constants which determine the upper bounds in Theorem~\ref{thm_for_mixture_families}.
We already have seen
\begin{align}
  \lambdabar \leq \frac{K}{\tau^2}
\end{align}
and
\begin{align}
  \Lambda
  &\leq
    \pair[\Big]{\frac{K}{\tau^2}}^K.
\end{align}
Then, we have an upper bound for the constant
\begin{align}
  C_J &= (\lambdabar^{1/2} + 2)^{K - 1} \Lambda^{1/2} \\
  &\leq
  \pair[\Big]{\frac{\sqrt{K}}{\tau} + 2}^{K - 1} \pair[\Big]{\frac{\sqrt{K}}{\tau}}^K.
\end{align}
Further, since 
\[
\min \{ \theta_i : \theta \in \Theta_\tau \} = \tau
\]
and 
\[
\max \{ \theta_i : \theta \in \Theta_\tau \} = 1-K\tau
\]
hold for all $i$,
we have $W_\Theta = 1 - (K+1)\tau$.
Hence, we have
\begin{align}
  C_K = 2^K (W_\Theta + 2a)^{K - 1}
  = 2^K (1 - (K + 1) \tau + 2a)^{K - 1}.
\end{align}
We also have an upper bound on $C_{J,K}$, defined as \eqref{definition_C_JK}:
\begin{align}
  C_{J, K}
  \leq
  \pair[\Big]{\frac{\sqrt{K}}{\tau}}^{K} \frac{C_K a}{\int_{\Theta} \abs{J(\theta)}^{1/2} d\theta}.
\end{align}
The constant $D_J$ in Assumption~\ref{assumption_D_J}, which determines $C_{\Theta}$, depends on $\set{q_i}$.
We cannot evaluate it generally.
If we have an evaluation of $D_J$ for given $\set{q_i}$,
we can determine $C_{\Theta}$ and then $r(n)$.

\section{Concluding Remarks}
We have developed a way of construction of two part codes which
have regret bounds asymptotically close to the minimax regret
for non-exponential families under Assumptions~\ref{assumption_cont_differentiability}--\ref{assumption_V2}.
The regret bounds are
given by Theorem~\ref{thm_improved_regret_bound}.
The regret bounds lead to the risk bounds of MDL estimators induced by the two part codes,
which are given by Corollary~\ref{cor_risk_bound_for_non_exponential_families}.
Our two part codes and MDL estimators allow us
to estimate probability densities in the non-exponential families which satisfy
Assumptions~\ref{assumption_cont_differentiability}--\ref{assumption_V2},
under guarantee of the risk.
Further, we have shown that
mixture families satisfy the assumptions we need and
have constructed such codes for mixture families.
To examine other families is a future work.


%


\ifCLASSOPTIONcaptionsoff
\newpage
\fi



%
\bibliographystyle{IEEEtranS}
\bibliography{references}   

\appendices

\section{Proof of Theorem 1}
\label{app1}

Since $\bar{d}_\lambda(p\|q)$ is increasing in $\lambda \in (0,1)$,
the proof for the case that
$\lambda = 1-\alpha^{-1}$ is sufficient.
Hence we assume it.
We have
\begin{align}\label{main_manipu_BC}
    &\E\Bigl( \bar{d}_\lambda(p^*\|\ddot{p}) -\frac{1}{n} \log\frac{p^*(X^n)}{\ptp(X^n)} \Bigr) \\
    = &
    \E\Bigl( \bar{d}_\lambda(p^*\|\ddot{p}) -\frac{1}{n}\log\frac{p^*(X^n)}{\ddot{p}(X^n)} 
    -\frac{\alpha L_n(\ddot{p})}{n}\Bigr)
    \\
    =&
    \frac{\alpha}{n}
    \E\Bigl( \frac{n \bar{d}_\lambda(p^*\|\ddot{p})}{\alpha} -\frac{1}{\alpha}\log\frac{p^*(X^n)}{\ddot{p}(X^n)} 
    - L_n(\ddot{p})\Bigr)
        \\
    =&
    \frac{\alpha}{n}
    \E\log \exp \Bigl( \frac{n \bar{d}_\lambda(p^*\|\ddot{p})}{\alpha} -\frac{1}{\alpha}\log\frac{p^*(X^n)}{\ddot{p}(X^n)} 
    - L_n(\ddot{p})\Bigr)
            \\
    \le &
    \frac{\alpha}{n}
    \log \E \exp \Bigl( \frac{n \bar{d}_\lambda(p^*\|\ddot{p})}{\alpha} -\frac{1}{\alpha}\log\frac{p^*(X^n)}{\ddot{p}(X^n)} 
    - L_n(\ddot{p})\Bigr)
    \\
        \le &
    \frac{\alpha}{n}
    \log \E \sum_{\bar{p}\in \ddPset}\exp \Bigl( \frac{n \bar{d}_\lambda(p^*\|\bar{p})}{\alpha} -\frac{1}{\alpha}\log\frac{p^*(X^n)}{\bar{p}(X^n)} 
    -L_n(\bar{p})\Bigr)
    \\
    = &
     \frac{\alpha}{n}
    \log \sum_{\bar{p}\in \ddPset} \E \exp \Bigl( \frac{n \bar{d}_\lambda(p^*\|\bar{p})}{\alpha} 
    -\frac{1}{\alpha}\log\frac{p^*(X^n)}{\bar{p}(X^n)} 
    -L_n(\bar{p})\Bigr).
\end{align}
The first inequality follows from Jensen's inequality.
The summation in the last line is
\begin{align}
  \sum_{\bar{p}\in \ddPset} \E \exp \Bigl( \frac{n \bar{d}_\lambda(p^*\|\bar{p})}{\alpha} 
    -\frac{1}{\alpha}\log\frac{p^*(x^n)}{\bar{p}(x^n)} 
    -L_n(\bar{p})\Bigr)
    =
 \sum_{\bar{p}\in \ddPset}  \exp \Bigl( \frac{n \bar{d}_\lambda(p^*\|\bar{p})}{\alpha} 
    \Bigr)  e^{-L_n(\bar{p})}
    \E\Bigl(\frac{\bar{p}(x^n)}{p^*(x^n)} \Bigr)^{1/\alpha},
\end{align}
where
\begin{align}
    \E\Bigl(\frac{\bar{p}(X^n)}{p^*(X^n)} \Bigr)^{1/\alpha}
=
\E\Bigl(\prod_t\frac{\bar{p}(X_t)}{p^*(X_t)} \Bigr)^{1/\alpha}
=\prod_t
\E\Bigl(\frac{\bar{p}(X_t)}{p^*(X_t)} \Bigr)^{1-\lambda}
=\E\Bigl(\frac{\bar{p}(X)}{p^*(X)} \Bigr)^{n(1-\lambda)}.
\end{align}
Further, recalling the definition of R\'enyi divergence
\begin{align}
    \exp \Bigl( \frac{n \bar{d}_\lambda(p^*\|\bar{p})}{\alpha} 
    \Bigr)
    & =
    \exp (n (1-\lambda)\bar{d}_\lambda(p^*\|\bar{p}))\\
    & =
    \exp\Bigl(n (1-\lambda) \Bigl(-\frac{1}{1-\lambda} \log \E \Bigl(\frac{\bar{p}(X)}{p^*(X)})\Bigr)^{1-\lambda}\Bigr)\Bigr)\\
    &=\E\Bigl(\frac{\bar{p}(X)}{p^*(X)} \Bigr)^{-n(1-\lambda)}.
\end{align}
Hence
\begin{align}\label{proof_of_BC}
  \sum_{\bar{p}\in \ddPset} \E \exp \Bigl( \frac{n \bar{d}_\lambda(p^*\|\bar{p})}{\alpha} 
    -\frac{1}{\alpha}\log\frac{p^*(X^n)}{\bar{p}(X^n)} 
    -L_n(\bar{p})\Bigr)
=\sum_{\bar{p}\in \ddPset}e^{-L_n(\bar{p})}\le 1,
\end{align}
which implies
\begin{align}
       \E\Bigl( \bar{d}_\lambda(p^*\|\ddot{p}) -\frac{1}{n} \log\frac{p^*(X^n)}{\ptp(X^n)} \Bigr) 
    \le 0.
\end{align}
This is the first inequality in the theorem.
The second inequality is derived as follows
\begin{align}
    \frac{1}{n} \RED^{(n)} (p^*, \ptp)
    = &
    \frac{1}{n} \E \Bigl( \log\frac{p^*(X^n)}{\ddot{p}(X^n)e^{-\alpha L_n(\ddot{p})}}  \Bigr)\\
    = &
    \frac{1}{n} \E \min_{\bar{p}\in \ddotPset}\Bigl( \log\frac{p^*(X^n)}{\bar{p}(X^n)e^{-\alpha L_n(\bar{p})}}  \Bigr)\\
        \le &
    \frac{1}{n} \min_{\bar{p}\in \ddotPset}\E \Bigl( \log\frac{p^*(X^n)}{\bar{p}(X^n)e^{-\alpha L_n(\bar{p})}}  \Bigr)\\
    = &
    \frac{1}{n}\min_{\bar{p}\in \ddotPset}(n D(p^*\|\bar{p}) +\alpha L_n(\bar{p}) )\\
      = &
    \min_{\bar{p}\in \ddotPset}\Bigl( D(p^*\|\bar{p}) +\frac{\alpha L_n(\bar{p})}{n} \Bigr).
\end{align}

\section{Proof of Theorem 2}
\label{app2}

Note that
\eqref{proof_of_BC} implies that
the expectation in the fifth line of \eqref{main_manipu_BC}
is not greater than 1.
Hence, by Markov's inequality,
we have
\begin{align}
    \Pr\Bigl\{ 
    \exp \Bigl( \frac{n \bar{d}_\lambda(p^*\|\ddot{p})}{\alpha} -\frac{1}{\alpha}\log\frac{p^*(X^n)}{\ddot{p}(X^n)} 
    - L_n(\ddot{p})\Bigr) > e^{nb\alpha^{-1}}
    \Bigr\}
    > e^{-n\alpha^{-1}b},
\end{align}
which is
\begin{align}
    \Pr\Bigl\{ 
    \bar{d}_\lambda(p^*\|\ddot{p})     
    >
    \frac{1}{n}
    \log\frac{p^*(X^n)}{\ptp(X^n)} +
    b
    \Bigr\}
    \le e^{-n\alpha^{-1}b}.
\end{align}

\end{document}

%% file: macro.tex
\usepackage{color}

\newcommand{\bluenote}[1]{\textcolor{blue}{#1}}
\newcommand{\greennote}[1]{\textcolor{green}{#1}}
\newcommand{\magentanote}[1]{\textcolor{magenta}{#1}}

\DeclarePairedDelimiter\set{ \{ }{ \} }

\DeclarePairedDelimiter\pair{(}{)}
\DeclarePairedDelimiter\abs{\lvert}{\rvert}
\DeclarePairedDelimiter\dabs{\lVert}{\rVert}

\DeclarePairedDelimiter\interval{[}{]}
\DeclarePairedDelimiter\partv{.}{|}

\newcommand{\E}{{\mathrm E}}

\newcommand{\Tr}{\mathrm{Tr}}

\newcommand{\argmax}{\arg \max}
\newcommand{\argmin}{\arg \min}

\newcommand{\Pset}{\mathcal{P}}
\newcommand{\Xset}{\mathcal{X}}
\newcommand{\Mset}{\mathcal{M}}

\newcommand{\Rset}{\mathbb{R}}

\newcommand{\Zset}{\mathbb{Z}}
\newcommand{\Gset}{\mathcal{G}_n}

\newcommand{\deltacoef}{g}

\newcommand{\ddothat}[1]{\ddot{\hat{{#1}}}}
\newcommand{\REG}{\mathrm {REG}}
\newcommand{\RED}{\mathrm {RED}}

\newcommand{\twopart}{{\rm tp}}
\newcommand{\Ltwopart}{L_{\twopart}}
\newcommand{\ptwopart}{p_{\twopart}}
\newcommand{\ptwopartbar}{\bar{p}_{\twopart}}
\newcommand{\pbar}{\bar{p}}
\newcommand{\dbar}{\bar{d}}
\newcommand{\Lbar}{\bar{L}}
\newcommand{\lambdabar}{\bar{\lambda}}

\newcommand{\Jhat}{\hat{J}}
\newcommand{\thetahat}{\hat{\theta}}
\newcommand{\thetaddothat}{\ddothat{\theta}}
\newcommand{\thetaddot}{\ddot{\theta}}
\newcommand{\Thetaddot}{\ddot{\Theta}}
\newcommand{\xiddot}{\ddot{\xi}}

\newcommand*{\interior}[1]{#1^\circ}
\newcommand*{\boundary}[1]{\partial #1}
\newcommand{\ddotPset}{\ddot{\Pset}}

\newcommand{\Scell}{\mathfrak{S}}
\newcommand{\Indi}{\mathbbm{1}}

\newcommand{\Ltp}{\Ltwopart}
\newcommand{\ptp}{\ptwopart}
\newcommand{\ptpbar}{\ptwopartbar}

\newcommand{\ddotTheta}{\Thetaddot}

\newcommand{\thetaddh}{\thetaddothat}
\newcommand{\ddPset}{\ddotPset}
\newcommand{\del}{\nabla}
\newcommand{\Del}{\del}

%% file: src.bbl
\begin{thebibliography}{10}
\providecommand{\url}[1]{#1}
\csname url@samestyle\endcsname
\providecommand{\newblock}{\relax}
\providecommand{\bibinfo}[2]{#2}
\providecommand{\BIBentrySTDinterwordspacing}{\spaceskip=0pt\relax}
\providecommand{\BIBentryALTinterwordstretchfactor}{4}
\providecommand{\BIBentryALTinterwordspacing}{\spaceskip=\fontdimen2\font plus
\BIBentryALTinterwordstretchfactor\fontdimen3\font minus
  \fontdimen4\font\relax}
\providecommand{\BIBforeignlanguage}[2]{{%
\expandafter\ifx\csname l@#1\endcsname\relax
\typeout{** WARNING: IEEEtranS.bst: No hyphenation pattern has been}%
\typeout{** loaded for the language `#1'. Using the pattern for}%
\typeout{** the default language instead.}%
\else
\language=\csname l@#1\endcsname
\fi
#2}}
\providecommand{\BIBdecl}{\relax}
\BIBdecl

\bibitem{Amari_Nagaoka}
S.~Amari and H.~Nagaoka, \emph{\textit{Methods of information geometry}}.\hskip
  1em plus 0.5em minus 0.4em\relax American Mathematical Soc., 2007, vol. 191.

\bibitem{Barron_Cover}
A.~R. Barron and T.~M. Cover, ``Minimum complexity density estimation,''
  \emph{\textit{IEEE Trans. on Inf. Theory}}, vol.~37, no.~4, pp. 1034--1054,
  1991.

\bibitem{barron1985logically}
A.~R. Barron, ``Logically smooth density estimation,'' Ph.D. dissertation,
  Stanford University, 1985.

\bibitem{Barron_etal2008}
A.~R. Barron, C.~Huang, J.~Q. Li, and X.~Luo, ``Mdl, penalized likelihood, and
  statistical risk,'' in \emph{2008 IEEE Information Theory Workshop}, 2008,
  pp. 247--257.

\bibitem{Chatterjee_Barron}
S.~Chatterjee and A.~R. Barron, ``Information theory of penalized likelihoods
  and its statistical implications,'' \emph{arXiv:1401.6714v2}, pp. 1--17,
  2014.

\bibitem{Grunwald}
P.~D. Gr{\"u}nwald, \emph{\textit{The minimum description length
  principle}}.\hskip 1em plus 0.5em minus 0.4em\relax MIT press, 2007.

\bibitem{Li_estimation_mixture}
Q.~J. Li, ``Estimation of mixture models,'' Ph.D. dissertation, Yale
  University, 1999.

\bibitem{Renyi}
A.~R\'enyi, ``On measures of entropy and information,'' in \emph{\textit{Proc.
  of the Fourth Berkeley Symp. on Math. Stat. and Prob.}}, vol.~1, 1961, pp.
  547--561.

\bibitem{Rissanen_1978}
J.~Rissanen, ``Modeling by shortest data description,''
  \emph{\textit{Automatica}}, vol.~14, no.~5, pp. 465--471, 1978.

\bibitem{Rissanen_1996}
------, ``Fisher information and stochastic complexity,'' \emph{\textit{IEEE
  Trans. on Inf. Theory}}, vol.~42, no.~1, pp. 40--47, 1996.

\bibitem{Shtarkov}
Y.~M. Shtar'kov, ``Universal sequential coding of single messages,''
  \emph{\textit{Problemy Peredachi Informatsii}}, vol.~23, no.~3, pp. 3--17,
  1987.

\bibitem{Takeuchi_Barron_1998}
J.~Takeuchi and A.~R. Barron, ``Asymptotically minimax regret by {B}ayes
  mixtures,'' in \emph{\textit{Proc. IEEE International Symp. on Inf. Theory}},
  1998, p. 318.

\bibitem{Takeuchi_Barron_2013}
------, ``Asymptotically minimax regret by {B}ayes mixtures for non-exponential
  families,'' in \emph{\textit{2013 IEEE Information Theory Workshop (ITW)}},
  2013, pp. 1--5.

\bibitem{Takeuchi_Barron}
------, ``Asymptotically minimax regret for models with hidden variables,'' in
  \emph{\textit{Proc. IEEE International Symp. on Inf. Theory}}, 2014, pp.
  3037--3041.

\bibitem{Takeuchi_Barron_2014}
------, ``Stochastic complexity for tree models,'' in \emph{2014 IEEE
  Information Theory Workshop (ITW 2014)}, 2014, pp. 222--226.

\bibitem{Zhang_convergence_MDL}
T.~Zhang, ``On the convergence of mdl density estimation,'' in
  \emph{International Conference on Computational Learning Theory}.\hskip 1em
  plus 0.5em minus 0.4em\relax Springer, 2004, pp. 315--330.

\bibitem{Zhang_2006}
------, ``From $\epsilon$-entropy to kl-entropy: Analysis of minimum
  information complexity density estimation,'' vol.~34, no.~5, 2006, pp.
  2180--2210.

\end{thebibliography}
